\documentstyle[epsfig,prc,preprint,aps]{revtex}
\begin{document}
\draft

\tightenlines

\begin{title}
{$^{\;\;\, 6}_{\Lambda\Lambda}$He as a ${\bf \Lambda\Lambda}$ Interaction 
Constraint}
\end{title}

\author{S.\ B.\ Carr and I.\ R.\ Afnan}
\address{Department of Physics, Faculty of Science and Engineering,
         Flinders University of South Australia,
         Bedford Park, SA 5042, Australia} 
\author{B.\ F.\ Gibson}
\address{Theoretical Division, Los Alamos National Laboratory, \\
         Los Alamos, NM 87545, USA}

\date{\today}

\maketitle

\begin{abstract}
The Nijmegen OBE potential D is $SU(3)$ rotated to model the strangeness -2 
sector of the baryon-baryon force.  Soft core repulsion is introduced to 
regularize the singular nature of the OBE functions.  The strength of the 
$^1$S$_0$ $t=0$ interaction is adjusted to model three 
scenarios: (1) a bound state, (2) a narrow virtual, or antibound, state, 
and (3) an unbound state in which the force is weakly attractive.  Using a 
separable approximation to these potentials, the binding energy of 
$^{\;\;\, 6}_{\Lambda\Lambda}$He is calculated in an $\Lambda\Lambda\alpha$ 
model.  The resultant binding energies suggest that the strength of the 
$^1$S$_0$ $t=0$ $\Lambda\Lambda$ and the $nn$ interactions should be similar, 
if the coupling between the $\Lambda\Lambda$ and $\Xi N$ channels is taken 
into consideration.
\end{abstract}

\newpage
\section{Introduction}\label{sec1}

The doubly strange $\Lambda \Lambda$ and $\Xi$ hypernuclei provide the
primary data that address the question of the properties of the $S=-2$
(strangeness $-2$) baryon-baryon interaction.  Direct two-body scattering 
is impractical.  In contrast, the data base for $S=0 \; \; NN$ scattering 
is relatively complete.  Furthermore, although $YN$ scattering data in the 
$S=-1$ sector are sparse, one does have differential and total cross section 
measurements to constrain theoretical models.  Thus, there exists a keen 
interest in the reported $\Lambda \Lambda$ \cite{dan63,pro66,aok91,dov91} 
and $\Xi$ \cite{wil59,bar63,bho63,bec68,cat69,mon79} hypernuclear events 
and in proposals to make new measurements at the Brookhaven AGS.

Our ultimate goal is to investigate the possibility of using the 
neutron spectrum from the $\Xi^- + d \to \Lambda + \Lambda + n$ 
reaction as a means to explore $\Lambda \Lambda$ scattering in the 
final state.  The analogous proton spectrum from the low energy
$n + d \to n + n + p$ reaction \cite{tor96} exhibits a sizable 
final-state interaction peak which is governed by the $nn$ scattering 
length and effective range.  To carry out that investigation, we must 
first model the $\Lambda \Lambda$--$\Xi N$ interaction using the 
available data on $\Lambda \Lambda$ and $\Xi$ hypernuclei as constraints.
The hypernucleus $^{\;\;\, 6}_{\Lambda \Lambda}$He becomes our laboratory 
for this purpose.  

The single reported $^{\;\;\, 6}_{\Lambda \Lambda}$He event \cite{pro66}
is controversial \cite{dal89}.  However, the $\Lambda \Lambda$ separation 
energy
\begin{equation}
   B_{\Lambda \Lambda} = B(^{\;\;\, 6}_{\Lambda \Lambda}He) 
              - B(^4He) 
     \simeq 10.92 \; \pm \; 0.6 {\rm MeV}       \label{eq:1}
\end{equation}
yields a value for the $\Lambda \Lambda$ interaction matrix element
\begin{equation}
  -\, \langle V_{\Lambda \Lambda} \rangle = B_{\Lambda \Lambda} -
               2 \times B_{\Lambda}(^5_{\Lambda}He)
   \simeq 4.7 \; {\rm MeV} \; ,                  \label{eq:2}
\end{equation}
which is consistent with the values of $4-5$ MeV extracted from
analysis of the heavier $\Lambda \Lambda$ hypernuclear events
\cite{dan63,aok91,dov91}.  It is this $\Lambda \Lambda$ separation 
energy that serves to constrain the freedom in our modeling of the 
$\Lambda \Lambda$--$\Xi N$ potential.

To minimize the number of parameters required in our strong interaction 
model, it is important to seek a unified model of the 
baryon-baryon force.  In the one-boson-exchange (OBE) model approach
\cite{swa71,nijD,nijF,nijSC,swa94,tim94}, $SU(3)$ symmetry is 
assumed to hold at the level of coupling constants and vertices.  
(Alternative $YN$ potential models have been produced by the J\"ulich
group \cite{jul89,hol92,jul92,reu94}.)  The $SU(3)$ symmetry is then 
broken by the use of physical masses for the baryons and mesons and 
by imposing phenomenological cutoff parameters to account for the 
short range properties of the forces.  With no $S=-2$ scattering data, 
we employ an $SU(3)$ rotation from the $S=0$ and $S=-1$ sectors to fix 
the $S=-2$ coupling constants.  Only the short range cutoff parameters 
remain to be determined.

The $S\neq0$ sector differs markedly in one aspect from the $S=0$
sector.  In the $S=0$ sector one includes $NN$--$\Delta N$ coupling 
between the octet and decuplet implicitly in the OBE approach. However,
the explicit inclusion of this coupling appears to play a relatively 
insignificant role in binding the $^3$H, $^3$He, and $^4$He 
few-nucleon systems.  In contrast, the $\Lambda N$--$\Sigma N$ coupling 
of different octet members (the mass difference is $\approx$77 MeV) is 
a major feature of the $S=-1 \; \; YN$ interaction, one which plays an 
important role in the obvious charge symmetry breaking exhibited by 
the $A=4$ isodoublet \cite{jur73,dav67,dav91,bam73,bej79} and the 
anomalously small binding of $^5_{\Lambda}$He \cite{jur73,dav67,dav91}.  
Therefore, one anticipates that the $\Lambda \Lambda$--$\Xi N$ coupling 
of different octet members (the mass difference is only $\approx$25 MeV) 
may play an even larger role in the $S=-2$ sector.

The strong $nn \; ^1$S$_0$ interaction (almost bound system) has 
a possible $S=-2$ sector analog in the $\Lambda\Lambda$--$\Xi N$ 
interaction.  That is, one might expect the $V_{\Lambda\Lambda}$ 
component of $V_{\Lambda\Lambda-\Xi N}$ to be a sizeable fraction of 
the strength of $V_{nn}$, if the \{27\} is the dominant $SU(3)$ 
representation \cite{dov90,dov92,dovpr}; $\Lambda\Lambda$--$\Xi N$
coupling would then enhance the strength of the full 
$\Lambda\Lambda$--$\Xi N$ interaction, possibly to something 
comparable to that of the $nn$ interaction.  
However, Hatree-Fock investigations \cite{gib69} imply that a 
$V_{\Lambda \Lambda}$ potential comparable in strength to that of the 
relatively weak $V_{\Lambda N}$ potential accounts for the binding energies 
of $^{\;\;\, 6}_{\Lambda\Lambda}$He and $^4_{\Lambda}$He.  That is, one 
finds
\begin{equation}
   \langle V_{\Lambda \Lambda} \rangle_{A=6} \simeq 
                   \langle V_{\Lambda N} \rangle_{A=4}   \label{eq:3}
\end{equation}
can account for the binding energy of both systems.  A similar 
conclusion was reached by Bodmer {\it et al.} \cite{bod84} and by 
Band\=o \cite{ban82}.

This would-be discrepancy between a strong $V_{\Lambda\Lambda-\Xi N}$ 
potential and the relatively weaker extracted value of 
$\langle V_{\Lambda \Lambda} \rangle_{A=6}$ can be understood in 
terms of the probable suppression of $\Lambda \Lambda$--$\Xi N$ 
coupling in the $A=6$ system.  Just as the $\Lambda N$--$\Sigma N$
coupling appears to be suppressed in $^5_{\Lambda}$He 
\cite{bod66,her67,gib72,gib72a,gal75,bod88} (the conversion 
of the $t=0 \; \; \Lambda$ into a $t=1 \; \; \Sigma$ requires a 
corresponding transition from the $t=0 \; \; \alpha$ core to an 
even-parity $t=1$ excitation high in the $\alpha$ spectrum
\cite{gib69,bod88,car91}), the $\Lambda \Lambda$--$\Xi N$ coupling is 
likely suppressed in $^{\;\;\, 6}_{\Lambda \Lambda}$He, because 
converting an $^1$S$_0 \; \, \Lambda\Lambda$ pair into a similar 
$\Xi N$ pair will require either excitation of the $\alpha$
core or placement of the fifth nucleon in the $\Xi N^5$ system in a 
$2s$ state in the shell model picture, due to the Pauli principle.
Therefore, we argue that $^{\;\;\, 6}_{\Lambda \Lambda}$He 
probes primarily the diagonal element $V_{\Lambda \Lambda}$ of
the $\Lambda \Lambda$--$\Xi N$ potential.  (In contrast
$^{\;\;\, 4}_{\Lambda \Lambda}$H would probe the full
$\Lambda \Lambda$--$\Xi N$ coupling and 
$^{\;\;\, 5}_{\Lambda \Lambda}$He would strongly enhance
the role of the $\Lambda \Lambda$--$\Xi N$ coupling because the 
$^4_{\Lambda}$He--$^4$He binding energy difference would
reduce dramatically the energy denominator in the transition.
The $^{\;\;\, 5}_{\Lambda \Lambda}$H system was addressed as
a three-body model by Myint and Akasishi \cite{Myi94}.)

We report here bound-state calculations in which we test primary 
features of a series of two-body $\Lambda \Lambda$--$\Xi N$ 
separable potentials within the realm of a three-body 
$\Lambda \Lambda \alpha$ model of 
$^{\;\;\, 6}_{\Lambda \Lambda}$He~\cite{bod88}.  We explore the 
effect of including explicit $\Lambda \Lambda$--$\Xi N$ coupling.  
In particular, we are able to find a model which produces an 
anti-bound state in the $\Lambda \Lambda$--$\Xi N$ system and 
also yields agreement with the experiment for 
$B_{\Lambda \Lambda}(^{\;\;\, 6}_{\Lambda \Lambda}$He).
In addition, we examine models in which the 
$\Lambda \Lambda$--$\Xi N$ system is bound and in which it is
weakly interacting.  One would like to include the $0^+$
excited state of the $\alpha$ particle in the calculation.
However, given the paucity of physical observables available
to constrain the $\Lambda\alpha$ (and therefore $\Lambda\alpha^*$)
interaction, we must neglect any explicit $\Lambda\Lambda\alpha^*$ 
coupling effects in this investigation.

In the next section we discuss the parameterizations of the 
$\Lambda \Lambda$ and $\Xi N$ interactions which we use as input.
In Sec.\ III we formulate the three-body bound-state model of
$^{\;\;\, 6}_{\Lambda \Lambda}$He.  Our numerical results are 
summarized and briefly discussed in Sec.\ IV.  Our conclusions
are collected in Sec.\ V.

\section{The Two-Body Input: the $\Lambda\Lambda$ and $\Xi N$ 
Interactions}\label{sec2}

Ideally, to model the baryon-baryon (BB) interaction we need a description 
that is consistent with the underlying fundamental theory of the 
strong interaction, QCD, and yet at the same time provides a practical 
framework for calculations.  An example of a QCD motivated model
is the  non-relativistic Quark Cluster  Model (QCM) 
\cite{oka80,fae82,oka84} which has been applied extensively to 
the BB system \cite{oka93,koi90,oka87}. While such models may give a good
representation of the short range interaction, they are incapable 
of describing the longer range aspects of the  potential.
At intermediate to long distances a description of the potential in terms 
of physical baryons and mesons is expected to dominate.  Thus for the $NN$ 
force, the One-Pion-Exchange (OPE) tail is observed to 
be the main component of the force in this region.  Extending this 
OPE potential to include heavier mesons that parameterize multi-pion 
exchanges, theorists since the sixties have constructed OBE models that 
account for the intermediate to long range components of the strong force.  
In this picture the short range component of the force is necessarily 
handled phenomenologically, either by the introduction of form factors 
for the meson-baryon vertices or by the introduction of short range 
cutoffs. The Nijmegen model D \cite{nijD} is just such a OBE model, 
one which had as its prime motivation a description of all BB systems 
up to the pion production threshold within the same theoretical 
framework. This was achieved by invoking the internal symmetry 
$SU(3)_{f}$; that was assumed to be valid at the level of the
baryon-baryon-meson coupling constants but to be dynamically broken by 
the inclusion of physical baryon and meson masses and by the inclusion 
of phenomenological short range cutoffs.  We have extended the model to 
the $S=-2$ sector by performing an $SU(3)$ rotation for the coupling 
constants \cite{dovpr,dov84} and introducing a smooth short range cutoff
to regularize the interaction as short distance.

\subsection{OBE potentials in coordinate space}\label{sec2.1}

Starting from a Lagrangian picture, one
can show that the basic OBE potential form will be a sum of terms 
involving central, spin-spin, tensor and more general terms involving 
the coupling of the spin and orbital angular momentum \cite{swa71}. 
With the restriction to s-wave interactions only ({\it i.e.}, $\ell$ = 0), 
these general terms disappear and the potential assumes the form  
\begin{equation}
V_{\mbox{\scriptsize OBE}}(r) = V_{c}(r) + V_{\sigma}(r)\, \mbox{\boldmath
$\sigma$}_{1}\cdot \mbox{\boldmath $\sigma$}_{2} 
+ V_{T}(r)\, \mbox{\boldmath $S$}_{12}  ,             \label{eq:4}
\end{equation}
where
\begin{equation}
 {\bf S}_{12} =3\,\mbox{\boldmath $\sigma$}_{1}\cdot 
{\bf\hat r}\ \mbox{\boldmath $\sigma$}_{2}\cdot {\bf\hat r}
-  \mbox{\boldmath $\sigma$}_{1}\cdot 
\mbox{\boldmath $\sigma$}_{2}\                        \label{eq:5}
\end{equation}
couples $\ell$ = 0 and $\ell$ = 2 states for spin-triplet configurations.
The two identical $\Lambda$ hyperons must be in an antisymmetric  
state.  Because the $\Lambda$ carries isospin zero, and because we  
consider only s-wave two-body interactions, the only spin state allowed 
is the s = 0 configuration. This channel has no OPE component and 
consequently no tensor force, so that the term involving 
$\mbox{\boldmath $S$}_{12}$ vanishes.  The same is also true for the 
$ ^{1}S_{0}$ $t$ = 0 and $t$ = 1 $\Xi N$ channels.  For the 
$^{3}S_{1}$ $t$ = 0 and $t$ = 1 $\Xi N$ channels, we can safely 
ignore the tensor force in a first calculation, due to the fact that 
$g_{\Xi \Xi \pi}$ is small \cite{dov84} and it is the OPE potential 
that is primarily responsible for the tensor force component of the 
potential.  Making this approximation for the $ ^{3}S_{1}$ $\Xi N$ 
interactions, our potentials take the form
\begin{equation}
V_{\mbox{\scriptsize OBE}}(r) = \left(\  V_{c}(r) + V_{\sigma}(r) \ 
\mbox{\boldmath $\sigma$}_{1} \cdot \mbox{\boldmath $\sigma$}_{2}\ \right) 
       \left( \begin{array}{c}
             1 \\ 
             \mbox{\boldmath $t$}_{1} \cdot \mbox{\boldmath $t$}_{2}
       \end{array} \right) \ ,                   \label{eq:6}
\end{equation}
where we include the isospin factor 
$\mbox{\boldmath $t$}_{1} \cdot \mbox{\boldmath $t$}_{2}$, 
for any isovector meson exchanges that contribute to 
$V_{\mbox{\scriptsize OBE}}(r)$.  For the $\Xi N$ states of total isospin 
$t $ = 0, 1 the isospin factor is normalized so that $\mbox{\boldmath 
$t$}_{1} \cdot \mbox{\boldmath $t$}_{2} = (-3, +1)$ respectively.

An important aspect of the $S=-2$ interaction is the coupling between 
the $\Lambda\Lambda$ and $\Xi N$ channels bearing the same quantum 
numbers.  The $\Lambda\Lambda$ system has a threshold mass of 
2$m_{\Lambda}$ = 2231.2~MeV, while the $\Xi N$ threshold lies only 
$\sim$ 25~MeV above this value at $m_{\Xi}$ + $m_{N}$ = 2257.01~MeV. 
The proximity of these thresholds implies a need to 
include in the model a mechanism that allows for conversion 
between these two-particle states.  Comparison of this system with 
the $\Lambda N$-$\Sigma N$ system \cite{gib94}, where coupling is 
seen to play an important role in the determination of the binding 
energies of light $\Lambda$ hypernuclei, supports this 
conjecture.  It seems natural that the form of the OBE potential 
itself should provide this coupling through the exchange of 
strange mesons, namely the $K$ and $K^{*}$.  The Nijmegen scheme 
for achieving this is described briefly in the Appendix, where 
we detail the OBE potential in the $S=-2$ channel. 

We must solve the Lippmann-Schwinger (LS) equation for the low 
energy scattering parameters for the relevant two-body systems.   
Rather than constructing the required potentials directly in 
momentum space, we first work in coordinate space where the 
physics is more transparent, and then we transform to momentum
space.  Until this point, we have not needed any phenomenology 
to specify the form of the interaction.  The procedure 
we have adopted from the Nijmegen group is consistent and based 
on sound physical arguments, although to what extent $SU(3)$ is a
valid symmetry \cite{dov90,dov92} (at least for relating coupling 
constants) remains an open question.  However, we must now resort 
to pure phenomenology, to the extent that we cannot determine the 
short range part ({\it i.e.}, $r < 0.5$ fm) of the interaction from a 
pure OBE description.  Because we believe that the BB force is 
repulsive at short distances, we introduce a short range cutoff of 
the form ${\cal C}\,e^{-{\cal M}r}/({\cal M}r)$.  Here ${\cal C}$ 
and $\cal M$ (the cutoff mass) are free parameters that determine 
the size of the inner repulsion. The advantage of this procedure over 
introducing a hard core radius ($r_{c}$) cutoff is that we can quite easily 
transform this potential to momentum space, without having to
integrate something that is infinite for $ r < r_{c}$.

This prescription is somewhat analogous to Reid's \cite{rei68}
construction of his $NN$ soft-core potential.  However, we have not 
demanded the sophistication of Reid, for the lack of experimental 
data.  We have freedom in choosing the parameters ${\cal C}$ 
and $\cal M$; actually $\cal M$ can be fixed, leaving us with 
but one adjustable parameter.  Therefore, we resort to
analogy with other two-body systems, and to predicting binding
energies and scattering observables of few-baryon systems, to
determine the parameter ${\cal C}$, and thus to constrain the short 
range physics.

\subsection{Soft-core potentials}\label{sec2.2}

If we examine the spin-isospin dependence of the $S=-2$ s-wave 
OBE potentials, we find that the $t$ = 0 $ ^{1}S_{0}$ channel (i.e. the
$\Lambda\Lambda$--$\Xi N$) is the most attractive.
With essentially no experimental information at our disposal 
for the other $\Xi N$ channels, we have chosen to model their short range 
repulsion such that the overall potentials are weakly attractive, 
with effective range parameters $a$ $\sim$-1.75 to -1.94~fm and 
$r$ $\sim$3.0 to 3.28~fm.  The medium-to-long range components of 
these potentials ({\it i.e.}, the OBE parts alone) are used as a guide 
as to which potential is the weakest and which the strongest.

The most interesting question to ask is how we should model the 
short range force in the $\Lambda \Lambda$-$\Xi N$ system.  We have already 
noted that the coupling between the $\Lambda \Lambda$ and $\Xi N$ channels 
is expected to play an important role.   On the other hand, the uncertain 
short range physics will perhaps play no less a role once we have 
specified its form.  With this in mind we have chosen a series of four 
soft-core potentials for the $\Lambda \Lambda$-$\Xi N$ system, each with 
different two-body scattering properties.  We construct model A to be 
weakly attractive, model B to support an antibound state in the 
$\Lambda\Lambda$ system and to be comparable in strength to the $^{1}S_{0}$ 
$nn$ interaction, and two models which support $\Lambda\Lambda$ 
bound states, models C1 and C2.  Model C1 supports a bound state of 
$\sim$0.7~MeV, while model C2 supports a bound state of $\sim$4.7~MeV. 
Both bound-state energies are measured with respect to the 
$\Lambda\Lambda$ threshold.

The potentials take the following basic form 
\begin{equation}
V^{(i)}_{\xi}(r) = -V_{0}^{(i)} \ \left[ \ 
  \frac{e^{-m_{i} r}}{m_{i} r} - {\cal C} \ \left(\frac{{\cal
  M}}{m_{i}} \right) \ \frac{e^{-{\cal M} r}}{{\cal
  M} r} \ \right]  \ \ \xi = c,\  \sigma\ ,        \label{eq:7}
\end{equation}
where $m_i$ is the mass of the exchanged mesons $(i)$. 
Here, $V_{0}^{(i)}$ symbolically represents the product of various 
coupling constants and mass factors that characterize the OBE potential 
form (see the Appendix). We choose to fix $\cal M$ = 2500~MeV and 
vary ${\cal C}$ to produce the required balance between inner range 
repulsion and outer range attraction, to construct potentials with 
the various two-body properties enumerated above. Depending on our 
choice for the parameter ${\cal C}$ (see Table~\ref{table1})  
we generate potentials A, B, C1 and C2.  

For some mesons the OBE contribution is repulsive, ({\it i.e.} $V_0^{(i)}<0$
in Eq.~(\ref{eq:7})), and to guarantee that the cutoff is repulsive, 
we need to take ${\cal C}<0$ for these mesons.

\subsection{Momentum space potentials}\label{sec2.3}

In momentum space we construct the s-wave spin-isospin dependent 
potential as the $\ell=0$ component of the partial wave expanded 
two-body potential
\begin{equation}
V_{\gamma \beta}({\bf p},{\bf p}^{\prime}) = \sum_{n, l, m_{s}, m_{t}}
  \langle\hat{p}|l n m_{s} m_{t}\rangle\ V^{n l}_{\gamma \beta}(p,
  p^{\prime})\ \langle m_{t} m_{s} n l|\hat{p}\rangle \ ,\label{eq:8}
\end{equation}
where n = \{s,t\} is the total spin and isospin of the two-body state.  
The subscripts $\gamma$ and $\beta$ label the two-body channels, 
{\it i.e.} $\gamma,\beta$ = ($\Lambda$ or $\Xi$). The $s$-wave momentum space 
potential,  $V^{n 0}_{\gamma \beta}(p,p^{\prime})$ $\equiv$ 
$V^{n}_{\gamma \beta}(p, p^{\prime})$, is now given in terms of the 
corresponding coordinate space potential by the Bessel transform 
\begin{equation}
V^{n}_{\gamma \beta}(p, p^{\prime}) = \frac{2}{\pi} \int_{0}^{\infty} dr
  r^{2} j_{0}(p r )\ V^{n}_{\gamma \beta}(r)\  
  j_{0}(p^{\prime} r ) \ ,                         \label{eq:9}
\end{equation}
where we have chosen a $\delta$-function normalization for the plane waves.

We interpret $V^{n}_{\gamma \beta}(r)$ as the sum of all the meson
exchanges contributing to a given two-body state of definite total spin
and isospin in coordinate space, with a similar meaning for
$V^{n}_{\gamma \beta}(p, p^{\prime})$.  The explicit form for our
soft-core potentials in momentum space for a given exchanged meson ($i$) is 
\begin{equation}
V^{(i)}_{\xi}(p,p^{\prime}) = - \frac{V^{(i)}_{0}}{2 \pi m_{i} p
  p^{\prime}} \ \left[ \ \ln\left(\frac{a+1}{a-1}\right) 
  - {\cal C} \ \ln\left(\frac{b+1}{b-1} \right) \ \right]  \ ,\   
\xi = c,\ \sigma\ ,                                \label{eq:10}
\end{equation}
where
\begin{equation}
a = \frac{m_{i}^{2} + p^{2} + p^{\prime 2}}{2 p p^{\prime}}
\qquad\mbox{and}\qquad
 b = \frac{{\cal M}^{2} + p^{2} 
    + p^{\prime 2}}{2 p p^{\prime}}                \label{eq:11}
\end{equation}
and ${\cal C}$ and $\cal M$ are the original parameters introduced in coordinate
space.  Explicit forms for the OBE potentials in p-space, for a given
exchanged meson $i$, are obtained by replacing $\phi(m_{i} r)$ in the
Appendix with  
\begin{equation} 
\phi(p, p^{\prime};m_{i}) = \frac{1}{2\pi m_{i} p p^{\prime} }\ 
  \left[ \ \ln\left(\frac{a+1}{a-1}\right) 
  - {\cal C} \ \ln\left(\frac{b+1}{b-1} \right) \ \right] 
  \ .                                              \label{eq:12}
\end{equation} 

\subsection{Potential parameters}\label{sec2.4}

In momentum space, we can solve directly for the scattering amplitude
($T$-matrix) using standard $K$-matrix methods and Gauss-Legendre quadratures.
The different thresholds coming from the rest mass difference between the 
$\Lambda\Lambda$ and the $\Xi N$ channels resides in the diagonal two-body 
Green's function 
\begin{equation}
\left[\mbox{\boldmath $G_{0}$}\right]_{\gamma\beta} 
        = \delta_{\gamma\beta}\ \left[\ E - M_{\gamma} 
        - \frac{p^2}{2 \mu_{\gamma}}\ \right]^{-1}\ ,\label{eq:13}
\end{equation}
where $M_\gamma$ and $\mu_\gamma$ are the total and reduced mass 
in the channel $\gamma$.

The scattering length in a given channel $a_\gamma$ and effective range 
parameter in that channel $r_\gamma$, for our potentials,
are related to the diagonal elements of the $T$-matrix and the phase
shifts by the relation \cite{afn93}:
\begin{equation} 
-\frac{1}{a_{\gamma}} + \frac{1}{2} p_{\gamma}^{2} r_{\gamma}\ 
   =\ p_{\gamma} \cot \delta_\gamma \ 
   =\  -\, \frac{1 - i \pi \mu_{\gamma} p_{\gamma} T_{\gamma \gamma}}
    {\pi \mu_{\gamma} T_{\gamma \gamma}}\ .        \label{eq:14} 
\end{equation}
In this case the scattering length and effective range for the $\Xi N$ channel are 
complex due to the presence of the other open channel, which is why the
$T$-matrix $T_{\Xi \Xi}$ is not real at the $\Xi N$ threshold.
These parameters, along with the two-body binding energies of soft core
potentials C1 and C2, are presented in Tables~\ref{table2} 
and \ref{table3} for each of our two-body interactions.

\subsection{Separable potentials}\label{sec2.5} 

As is well known, if we assume a separable form for the two-body potentials, 
then the analysis of the three-body problem is greatly simplified.  For this 
reason we construct separable approximations to our soft-core potentials, 
by adjusting the parameters of the separable potentials to reproduce the 
scattering lengths and effective ranges shown in Tables~\ref{table2} 
and \ref{table3}. This we do in each spin-isospin channel, and in 
the case of the $\Lambda\Lambda$-$\Xi N$ s = 0, t = 0 channel we include 
the coupling phenomenologically by generalizing the form of our separable 
potential through the inclusion of the off-diagonal coupling strength 
$C_{\Lambda \Xi}$.  Because we consider only s-wave interactions, our 
separable potentials take the form 
\begin{equation} 
V^{n}_{\gamma \beta}(p,p^{\prime}) = g^{n}_{\gamma}(p)\  
  C^{n}_{\gamma\beta}\ g^{n}_{\beta}(p^{\prime})\ .      \label{eq:15}
\end{equation} 
We choose Yamaguchi form factors \cite{yam54} 
\begin{equation} 
g^{n}_{\gamma}(p) = ( p^{2} + \beta^{n^{2}}_{\gamma})^{-1}\ ,\label{eq:16}
\end{equation} 
which are given in terms of the range parameters $\beta^{n}_{\gamma}$.  The
sign of the coupling strength $C^{n}_{\Lambda \Xi}$ is determined by matching 
the sign of the $V_{\Lambda \Xi}$ matrix element for the separable 
potential with that of the soft-core potential.

For the single-channel potentials we can extract both the strength and
range parameters using analytic expressions:
\begin{equation} 
p_\gamma \cot \delta_\gamma  =  -\frac{\beta_\gamma}{2}\, \left( 1 + 
               \frac{2\beta_\gamma^3}{C_\gamma \pi\mu_\gamma}\right) 
               + \frac{p_\gamma^2}{2\beta_\gamma} 
\left( 1 - \frac{4 \beta_\gamma^3}{C_\gamma \pi\mu_\gamma}\right) 
  =  -\frac{1}{a_\gamma} + \frac{1}{2}r_\gamma p_\gamma^2 \ ,
\end{equation} 
where $\gamma$ is the channel label.
In the case of the $\Lambda\Lambda$-$\Xi N$ coupled-channels potentials, we 
determined the strength and range parameters by searching a five dimensional 
parameter space (for $C_{\Lambda \Lambda}$, $C_{\Lambda \Xi}$, $C_{\Xi N}$, 
$\beta_{\Lambda}$ and $\beta_{\Xi}$) for the optimum $\chi^{2}$ fit to the 
effective range parameters in both the $\Lambda\Lambda$ and $\Xi N$ channels.
The calculations were performed for different values of momenta less than
0.1 $fm^{-1}$, to insure that the effective range parameters did not
depend upon the choice of momenta.

The resulting separable potential parameters ({\it i.e.} the strength and 
range parameters) can be found in Tables~\ref{table4} and 
\ref{table5}. For comparison with the soft-core potential parameters, 
we also collect in Table~\ref{table6} the scattering lengths 
and effective ranges for the separable potentials in the 
$\Lambda\Lambda$-$\Xi N$ coupled channel.

\section{The ${\bf \Lambda\Lambda\alpha}$ Three-Body Model}\label{sec3}

The fact that $^4$He is tightly bound suggests that we may model 
$_{\Lambda\Lambda}^{\;\;\, 6}$He as a $\Lambda\Lambda\alpha$--$\Xi N\alpha$ 
system and thus neglect the possible excitation of the $^4$He core.  This 
puts some limitation on our model in terms of the role of the Pauli
principle between the nucleon and the $\alpha$ in the $\Xi N\alpha$ channel,
and therefore the role of the coupling between the $\Lambda\Lambda$--$\Xi N$
coupled channels.

In a simple shell model, the $\alpha$ core may be considered as four nucleons
in  the $1s$ shell with total quantum numbers $ J^{\pi} = 0^{+}$, $T = 0$. The
Pauli principle will require that a fifth nucleon be placed in either the
$p$ or $s-d$ shell, with the latter being suppressed by the fact that it
requires an additional $\hbar\omega$ in energy. In our three-body model this
Pauli effect can be implemented by either requiring that the $s$-wave $N\alpha$
potential have its one Pauli forbidden bound state projected out, or 
by assuming that the $s$-wave interaction is repulsive, as has been done 
with success in bound-state calculations of $^6$Li \cite{leh78,leh74,esk92}. 
In the present investigation we use a repulsive $N\alpha$ $s$-wave potential.

The addition of the strangeness degree of freedom to this picture requires 
that we treat light hypernuclei as bound states of a nuclear 
core plus $n$ hyperons.  By treating the nucleon and the hyperon as
distinguishable baryons,  we need not require that the hyperon and the nucleon
core satisfy any Pauli principle.  This makes the treatment of 
$^5_{\Lambda}$He and $^{\;\;\, 6}_{\Lambda\Lambda}$He as 
$\Lambda\alpha$ and $\Lambda\Lambda\alpha$ bound states particularly 
attractive.  In fact, we may use the experimental $\Lambda\alpha$ 
separation energy of $B_{\Lambda}( ^{5}_{\Lambda} \mbox{He}) \sim$ 3.1~MeV  
to constrain our $\Lambda\alpha$ potentials.

A major feature of the $\Lambda\Lambda$-$\Xi N$ interaction constructed in 
the previous section is the coupling between the $\Lambda\Lambda$ and 
$\Xi N$ channels in the $ ^{1}S_{0}$, $t$ = 0 partial wave.  As noted 
above, one might expect the effective $\Lambda\Lambda$ matrix element 
in hypernuclei ({\it i.e.} $-\langle V_{\Lambda\Lambda}\rangle$ ) 
to be comparable to that of the $nn$ 
interaction which is $-\langle V_{n n}\rangle \simeq 6-7$~MeV \cite{dov94}.  
We argue that in $^{\;\;\, 6}_{\Lambda\Lambda}$He the coupling in a realistic 
$\Lambda\Lambda$-$\Xi N$ free space interaction is strongly suppressed, 
because the conversion $\Lambda\Lambda\rightarrow\Xi N$ is Pauli blocked 
due to the fact that the fifth nucleon in $\Xi N\alpha$ should be in an 
antisymmetric state relative to the nucleons in the $\alpha$ particle. 
As a result, the diagonal $\Lambda\Lambda$ element of the interaction gives 
the major contribution to the binding energy.  The Pauli blocking results 
in a reduced value of $-\langle V_{\Lambda \Lambda} \rangle \simeq 4-5 $~MeV, 
in agreement with that extracted from the one experiment \cite{pro66}.  
In other words, we anticipate that we can explain the ``apparent'' conflict 
between a realistic $\Lambda\Lambda$-$\Xi N$ interaction and the anomalously 
small value for the extracted value of $-\langle V_{\Lambda \Lambda} \rangle$
in terms of the role played by $\Lambda\Lambda$-$\Xi N$ coupling.

To test this hypothesis we performed bound-state calculations involving: 
($i$) Only the $\Lambda\Lambda$ component of the amplitude due to the full 
$\Lambda\Lambda$-$\Xi N$ force.  ($ii$) The full $\Lambda\Lambda$-$\Xi
N$  interaction, which necessarily involves the introduction of the 
repulsive $S_{\frac{1}{2}}$ $N\alpha$ potential to model the Pauli blocking, 
the p-wave $\alpha N$ interactions and an $\Xi\alpha$ potential.  
($iii$) A single-channel effective $\Lambda\Lambda$ interaction designed 
to reproduce the scattering length and effective range parameters of the 
complete coupled-channel potential.  We have already indicated why we 
believe ($i$) to be valid.  Case ($ii$) provides a check on the effect of 
introducing a repulsive $S_{\frac{1}{2}}$ $\alpha N$\ interaction on 
the separation energy 
$B_{\Lambda \Lambda}(^{ \ \ \ \ \ \ \, 6}_{\Lambda\Lambda-\Xi N}$He) 
of the hypernucleus involving a $\Xi N$ component in the wave 
function.  Coupling between the three-body channels, 
$\Xi-(N\alpha)_{S_{\frac{1}{2}}}$ in a state of relative orbital 
angular momentum $\cal{L}$ = 0, and  $\Xi-(N\alpha)_{P_{\frac{1}{2}}}$ 
or $\Xi-(N\alpha)_{P_{\frac{3}{2}}}$ in a state of relative orbital 
angular momentum $\cal{L}$ = 1, is allowed and satisfies parity and 
angular momentum conservation.  Because the $P_{\frac{1}{2}}$ 
and the $P_{\frac{3}{2}}$ $N\alpha$ interactions are strong (both support
resonances), we should include  them in our bound-state calculation.   If the
above transitions between three-body channels are strong, then the existence of
an excited $^{ \ \ \ \ \ \ \, 6}_{\Lambda\Lambda - \Xi N}$He$^*$ nucleus
cannot  be ruled out.  Because the inclusion of these $N\alpha$ states (and 
also an $S_{\frac{1}{2}}$ $\Xi\alpha$ interaction) is not beyond our 
computational capability, exploring the effect on the binding energy 
is worth the effort.  However, the importance of these states involving 
p-wave $\alpha N$ interactions is limited on two accounts.   Firstly, 
the system must couple to the ``Pauli forbidden'',
$\Xi-(N \alpha)_{S_{\frac{1}{2}}}$ channel in a state of relative 
orbital angular momentum  $\cal{L}$ = 0, and secondly the  
probability for the transition between the three-body channels involving the 
$S_{\frac{1}{2}}$ $N\alpha$ two-body cluster and the $P_{\frac{1}{2}}$ 
and $P_{\frac{3}{2}}$ $N\alpha$ two-body clusters should be small.

\subsection{ The ${\bf \Lambda\alpha}$, ${\bf \Xi\alpha}$, and ${\bf N\alpha}$
potentials}\label{sec3.1}

There is little or no experimental data for the $\Lambda\alpha$ and 
$\Xi\alpha$ $S_{\frac{1}{2}}$ two-body systems.  For the $\Lambda\alpha$ 
interaction there does exist as a constraint the experimental binding energy 
for the ground state of $^{5}_{\Lambda}$He.  In that case we can also
use our knowledge of the $\Lambda N$ interaction and construct 
an effective $\Lambda\alpha$ potential by folding the $\Lambda N$ 
G-matrix \cite{bru54,gol57,day67} with the nucleon distribution for 
the $^4$He core \cite{yam85}.  Any free parameter can then be adjusted 
to reproduce $B_{\Lambda}(^{5}_{\Lambda}$He).

For the $\Lambda\alpha$ system, we take Yamamoto's and Band\={o}'s 
effective YNG potential in coordinate space \cite{yam85}, which was 
designed to parameterize their $\Lambda N$ $G$-matrix.   In constructing 
this $\Lambda N$ $G$-matrix for nuclear matter, they used the Nijmegen
model D $\Lambda N$-$\Sigma N$ coupled-channel potential as input. 
When applying this YNG potential to light hypernuclear systems such as 
$^5_{\Lambda}$He, they treated the Fermi momentum $k_{F}$ as an 
adjustable parameter.   We take $k_{F} = 0.932$~fm$^{-1}$, so that our 
$\Lambda\alpha$ interaction has a bound state of $\sim$ 3.1~MeV as 
observed experimentally. The central YNG potential has the form
\begin{equation}
  V_{\Lambda N}^{t l s}(r;k_{F}) = \sum_{i=1}^{3} w_{i}\ e^{-
  (\frac{r}{\beta_{i}})^{2}}                 \label{eq:18}
\end{equation}
with
\begin{equation}
  w_{i}(k_{F}) = a_{i} + b_{i}\, k_{F}  
                    + c_{i}\, k_{F}^{2}\ ,    \label{eq:19}
\end{equation}
where the parameters $a_{i}$, $b_{i}$, $c_{i}$, and the range
parameter $\beta_{i}$, can be found in Table~\ref{table7}. 

We are interested in the $t = 1/2$, $\ell = 0$, $s = 0$ and $s = 1$ channels, 
as the correct effective $\Lambda N$ potential to be folded with the 
nucleon distribution of $^4$He is a spin averaged sum of spin-triplet 
and spin-singlet interactions; {\it i.e.},
\begin{equation}
  V_{\Lambda N}^{eff}(r;k_{F}) 
       = \frac{3}{4}\,  V_{\Lambda N}^{\frac{1}{2} 0 1}(r;k_{F}) 
       + \frac{1}{4}\, V_{\Lambda N}^{\frac{1}{2} 0 
                                     0}(r;k_{F})\ . \label{eq:20}
\end{equation}
The $\Lambda\alpha$ potential is then given by 
\begin{equation}
  V_{\Lambda \alpha}(r) = \int d^{3} r^{\prime}\  
   \rho_{^4 \mbox{\scriptsize He}}(r^{\prime})\ 
  V_{\Lambda N}^{eff}(| \vec{r} - \vec{r}^{\ \prime} |) \label{eq:21}
\end{equation}
with 
\begin{equation}
  \rho_{^4 \mbox{\scriptsize He}}(r^{\prime}) 
= \frac{4}{\pi^{\frac{3}{2}}b^{3}} \ 
  e^{-(\frac{r^{\prime}}{b})^{2}} \ ,          \label{eq:22}
\end{equation} 
where $b^{2} = \frac{2}{3} ( r_{ch}^{2} - r_{p}^{2} )$ and where 
$r_{ch}$ = 1.71 and $r_{p}$ = 0.8 fm are the charge radius of 
$^4$He and the proton, respectively.

For comparison, we include the $\Lambda\alpha$ interaction of Kurihara {\it et
al.} \cite{kur85}, which is  parameterized as a two term Gaussian of the form
\begin{equation}
  V_{\Lambda \alpha}(r) = V_{R}\, \exp [ - (r/b_{R})^{2} ] - 
  V_{A}\, \exp [ - ( r/b_{A} )^{2} ]\  ,      \label{eq:23}
\end{equation}
where $V_{R}$ = 450.4~MeV, $V_{A}$ = 404.9~MeV, $b_{R}$ = 1.25~fm, 
and $b_{A}$ = 1.41 fm.  This potential is referred to as the Isle 
potential \cite{kur85}.  It predicts a value for the (weak) lifetime 
of $^{5}_{\Lambda}$He in good agreement with experiment, as well as 
reproducing the value for $B_{\Lambda}(^{5}_{\Lambda}$He) of 3.1~MeV.
In the absence of any data other than the $\Lambda$ separation energy
and the weak decay lifetime, it seemed prudent to restrict our
consideration to these potential models rather than treat the
$\Lambda\alpha$ effective range as a free parameter.

For the $\Xi\alpha$ system we have no experimental data at all.  In 
this instance we make use of the bound-state data we have for other 
light $\Xi$ hypernuclei \cite{wil59,cat69,mon63,bec68}, to construct a 
$\Xi$-nucleus potential that allows us to extrapolate to the 
$^{\; 5}_{\Xi}$He system.  In this approach we follow Dover and Gal 
\cite{dov83}; we take their $\Xi$-nucleus potential and apply it to 
$^{\; 5}_{\Xi}$He.  Their value for the $\Xi$-nucleus well depth is 
close to the theoretical prediction \cite{dov84} for such a quantity,
making use of Nijmegen model D for the $\Xi N$ interaction.
The $\Xi$-nucleus potential, referred to as potential DG, has a 
Woods-Saxon form 
\begin{equation}
  V_{\Xi\alpha}(r) = - V_{0 \Xi} ( 1 + e^{- (r-R)/a} )^{-1}\ ,\label{eq:24}
\end{equation}
where $R = r_{0} A^{\frac{1}{3}}$, $a = 0.65$~fm.   We take $r_{0} = 
1.1$~fm, which in turn implies that $V_{0 \Xi} = 24$~MeV.  Here $A$ 
is the mass number of the core nucleus under consideration, in our
case $A = 4$. 

We calculated the scattering length and effective range parameters 
for each of these potentials by transforming them to momentum space 
and solving the LS equation.   The corresponding binding energies 
are listed along with the low energy scattering parameters in 
Table~\ref{table8}.
Separable approximations to these potentials (SYNG, SIsle and SDG)
were constructed.  The parameters of these separable potentials were 
determined by demanding that the potentials reproduce the effective 
range parameters and binding energies given in Table~\ref{table8}. 
These separable potentials have the same momentum space representation 
introduced previously in Eqs.~(\ref{eq:15}) and (\ref{eq:16}) where
$\gamma=\beta=Y\alpha$ with $Y = \Lambda$ or $\Xi$.
The parameters of the potential $C_{Y \alpha}$ and $\beta_{Y \alpha}$ are 
determined from the effective range and scattering length through the 
following relations
\begin{equation}
  \beta_i = \frac{ 3 \pm \sqrt{ 9 - 16 \frac{r_i}{a_i}}}{r_i} 
   \quad;\quad\quad
  C_i = \frac{4 \beta_i^3}{\pi \mu_i( 1 - \beta_i r_i)} \label{eq:25}
\end{equation}
with $i=Y\alpha$ and $\mu_i = m_{Y} m_{\alpha}/(m_{Y} + m_{\alpha})$.

Separable potentials exist in the literature for the $N\alpha$
interactions in the $S_{\frac{1}{2}}$,  $P_{\frac{1}{2}}$ and  
$P_{\frac{3}{2}}$ channels.  They have been used extensively in 
applications to $\alpha$-d scattering and bound-state calculations 
({\it e.g.}, $^6$Li treated as an $NN\alpha$ system; see 
Ref.~\cite{esk92} and references cited therein).  We utilize the 
parameters given in Ref.~\cite{esk92}; we consider model A for the 
p-wave interactions only.  In momentum space these potentials 
assume the familiar form given in Eq.~(\ref{eq:15}) with 
$\gamma=\beta=N\alpha$ and $n=\ell$, the relative orbital angular momentum. 
The form factors for the various $N\alpha$ channels are given by
\begin{equation}
  g_{N \alpha}^{\ell}(p) = \frac{p^{\ell}}{[p^{2} + \beta_{N
  \alpha}^{\ell^{2}}]^{(\ell+1)}}\  .              \label{eq:26}
\end{equation}
In this case the parameters were adjusted to fit the $N\alpha$ scattering
data. The parameters of the $N\alpha$ potentials are given in 
Table~\ref{table9}.

\subsection{The three-body equations}\label{sec3.2}

One of the most convenient forms of the Faddeev equations for practical 
solution of both the scattering and bound-state problems is the form in 
which Alt, Grassberger, and Sandhas presented them, the AGS equations 
\cite{alt67}. 
These equations appear in operator form as follows:
\begin{equation}
  U_{\zeta \beta}(E) = \bar{\delta}_{\zeta \beta} G_{0}^{-1}(E) + 
  \sum_{\gamma} \bar{\delta}_{\zeta \gamma} T_{\gamma}(E) G_{0}(E) 
  U_{\gamma \beta}(E) ,                          \label{eq:27}
\end{equation}
where $\bar{\delta}_{\zeta \beta} = 1 - \delta_{\zeta \beta}$, $E$ 
is the total three-body energy, and $T_{\gamma}(E)$ is the two-body 
$T$-matrix for the interacting ($\zeta \beta$) pair in the three-body 
Hilbert space.  The three-particle free Green's function has the form
\begin{equation}
  G_{0}(E) = ( E - H_{0} )^{-1}\  ,             \label{eq:28}
\end{equation}
where $H_{0}$, the Hamiltonian for the three non-interacting particles, 
is given by
\begin{equation}
H_{0}  =  \sum_{j=1}^{3}\ \left[ m_{j} + \frac{k_{j}^{2}}{2 m_{j}}\right] 
       =  \sum_{j=1}^{3}\ m_{j} + \frac{p_{\zeta}^{2}}{2 \mu_{\zeta}} +
            \frac{q_{\zeta}^{2}}{2 \nu_{\zeta}}\ , \label{eq:29}
\end{equation}
where the second equality holds in the three-body center of mass. We have
included the rest  masses in our definition of $H_{0}$ to allow for the
correct incorporation of the different three particle thresholds when
different mass eigenstates  are coupled through the two-body transition
amplitude $T_{\gamma}$ \cite{afn89}.  Here $\vec{p}_{\zeta}$ stands for the
momentum of the pair ($\beta \gamma$), while $\vec{q}_{\zeta}$ designates
the momentum of the spectator $\zeta$ relative to the ($\beta \gamma$) pair.
The Jacobi momenta and the reduced masses $\mu_{\zeta}$ and 
$\nu_{\zeta}$ are standard. The AGS operators, when sandwiched between the
plane wave states $|\vec{q}_{\zeta} \vec{p}_{\zeta} \rangle$, represent the
transition  amplitudes for the process 
$\beta + (\gamma \zeta) \rightarrow \zeta + (\beta \gamma)$.
Using standard three-body techniques, one can obtain equations for the 
physical transition amplitudes 
\begin{equation}
  X_{\zeta \beta} = \langle \phi_{\zeta} |\,G_0\, U_{\zeta \beta}\, 
  G_0\,|\phi_{\beta} \rangle\  ,                 \label{eq:30}
\end{equation}
where $G_0|\phi_{\zeta} \rangle$ is the two-body state for the interacting 
pair ($\beta \gamma$).  For two-body interactions of separable form, the AGS
equations reduce to a standard set of coupled integral equation for the
amplitudes $X_{\zeta\beta}$.  Because our $\Lambda\Lambda\alpha$ has two
identical Fermions, the two $\Lambda$s, we need to antisymmetrize the
equations in return for a reduction in the number of coupled integral
equations. This antisymmetrization has been detailed for the $\pi
NN$~\cite{afn74} and $YNN$~\cite{afn89} systems to give a set of equations 
for the antisymmetric $\Lambda\Lambda\alpha$ amplitudes of the form
\begin{eqnarray}
  X_{\alpha \alpha} & = & Z_{\alpha \Lambda} \tau_{\Lambda} 
    X_{\Lambda \alpha} \nonumber \\
  X_{\Lambda \alpha} & = & Z_{\Lambda \alpha} \tau_{\alpha}
    X_{\alpha \alpha} + Z_{\Lambda \Lambda} \tau_{\Lambda}
     X_{\Lambda \alpha} \ .                     \label{eq:31}
\end{eqnarray}
If we label the two $\Lambda$s as 1 and 2, while the $\alpha$ as particle 3,
the Born terms in Eq.~(\ref{eq:31}) are given by
\begin{eqnarray}
  Z_{\alpha \Lambda}   & = & \sqrt{2} Z_{31}^{c} 
= \sqrt{2}\langle(12)3|G_0|(23)1\rangle \nonumber\\
  Z_{\beta\zeta} &=& Z_{\zeta\beta} \nonumber \\
  Z_{\Lambda \Lambda}  & = & - Z_{12}
= -(-1)^R\,\langle(23)1|G_0|((31)2\rangle\  . \label{eq:32}
\end{eqnarray}
Here the phase $R$ results from the exchange of particles 3 and 1 in the
ket~\cite{afn89}.

These equations for the $\Lambda\Lambda\alpha$ system can be extended to
include the coupling between $\Lambda\Lambda$ and $\Xi N$ two-body states,
to encompass the  contribution from the $\Xi N \alpha$ Hilbert space. 
Since we are only interested in the binding energy, we need to examine the 
spectrum of the kernel of the homogeneous equations, and in this case the
initial state can be any of a number of possible two cluster states. To write
an explicit form for our equation, we have assumed an initial state of
$(\Lambda\Lambda)\alpha$. In this case our coupled homogeneous integral
equations for the $\Lambda\Lambda\alpha$--$\Xi N\alpha$ system take the form
\begin{eqnarray}
\left( \begin{array}{c}                    
        X_{\alpha \alpha} \\
        X_{\Lambda \alpha} \\
        Y_{\alpha \alpha} \\
        Y_{\Xi \alpha} \\
        Y_{N \alpha}
       \end{array}  \right) 
= \left( \begin{array}{ccccc}
      0 & Z_{\alpha \Lambda} \tau_{\Lambda}^{\alpha \Lambda} & 0 & 0 & 0 \\
      Z_{\Lambda \alpha} \tau_{\alpha}^{\Lambda \Lambda} & Z_{\Lambda
\Lambda} \tau_{\Lambda}^{\alpha \Lambda} & Z_{\Lambda
\alpha} \tau_{\alpha}^{\Lambda \Xi} & 0 & 0 \\
      0 & 0 & 0 & Z_{\alpha \Xi} \tau_{\Xi}^{\alpha N} &
Z_{\alpha N} \tau_{N}^{\alpha \Xi} \\
      Z_{\Xi \alpha} \tau_{\alpha}^{\Xi \Lambda} & 0 & Z_{\Xi \alpha}
\tau_{\alpha}^{\Xi N} & 0 & Z_{\Xi N} \tau_{N}^{\alpha \Xi} \\
      Z_{N \alpha} \tau_{\alpha}^{\Xi N} & 0 & Z_{N \alpha}
\tau_{\alpha}^{\Xi N} & Z_{N \Xi} \tau_{\Xi}^{\alpha N} & 0
               \end{array}  \right)   
\times 
\left( \begin{array}{c}                     
        X_{\alpha \alpha} \\
        X_{\Lambda \alpha} \\
        Y_{\alpha \alpha} \\
        Y_{\Xi \alpha} \\
        Y_{N \alpha}
       \end{array}  \right) \ ,            \label{eq:33}
\end{eqnarray}
where the $Y$ give the transition amplitude  from the initial
$(\Lambda\Lambda)\alpha$ state to a final state of the $\Xi N\alpha$, 
{\it e.g.} $Y_{\Xi\alpha}$ is the amplitude for
$(N\alpha)\Xi\leftarrow(\Lambda\Lambda)\alpha$. In the $\Xi N\alpha$ channel
we have three distinguishable particles, and as a result, there are three 
$Y$ amplitudes, and no antisymmetry is required.

To obtain a better understanding of the structure of these coupled integral
equations as well as how the $\Lambda\Lambda$--$\Xi N$ coupling is introduced, 
we have written the kernel of the above coupled integral equations in the 
form of a product of the Born terms ${\bf Z}$ which is diagonal in the
$\Lambda\Lambda\alpha$--$\Xi N\alpha$ channels, and a quasi-particle propagator
{\boldmath $\tau$ }that has the coupling between the $\Lambda\Lambda$ and the $\Xi N$ channels,
{\it i.e.} ${\bf K} = {\bf Z}\times$\mbox{\boldmath $\tau$}, where
\begin{eqnarray}
{\bf Z} & = & \left( \begin{array}{ccccc}
                      0 & Z_{\alpha \Lambda} & 0 & 0 & 0 \\
                      Z_{\Lambda \alpha} & Z_{\Lambda \Lambda} & 0 & 0 & 0 \\
                      0 & 0 & 0 & Z_{\alpha \Xi} & Z_{\alpha N} \\
                      0 & 0 & Z_{\Xi \alpha} & 0 & Z_{\Xi N} \\
                      0 & 0 & Z_{N \alpha} & Z_{N \Xi} & 0
                      \end{array} \right)        \label{eq:34}
\end{eqnarray}
and
\begin{eqnarray} 
\mbox{\boldmath $\tau$} & = & \left( \begin{array}{ccccc}
                          \tau_{\alpha}^{\Lambda \Lambda} & 0 & 
  \tau_{\alpha}^{\Lambda \Xi} & 0 & 0 \\
                          0 & \tau_{\Lambda}^{\alpha \Lambda} & 0 & 0 & 0 \\ 
                          \tau_{\alpha}^{\Xi \Lambda} & 0 &
  \tau_{\alpha}^{\Xi N} & 0 & 0 \\
                          0 & 0 & 0 & \tau_{\Xi}^{\alpha N} & 0 \\
                          0 & 0 & 0 & 0 & \tau_{N}^{\alpha \Xi}
                                     \end{array} \right)\ .\label{eq:35}
\end{eqnarray}
If we now set the coupling between the $\Lambda\Lambda$ and $\Xi N$ channels
to zero, {\it i.e.}, $\tau_{\alpha}^{\Lambda\Xi}$ =
$\tau_{\alpha}^{\Xi\Lambda}$ = 0,  then the three-body matrix integral 
equation decouples into two sets of
coupled equations for the  $\Lambda\Lambda\alpha$ and $\Xi N\alpha$ Hilbert
spaces, respectively, with the former set of equations being just 
Eq.~(\ref{eq:31}).
The proper inclusion of the different thresholds, associated with the
mass difference between the $\Lambda\Lambda$ and $\Xi N$ two-body systems, is 
accounted for by measuring all two-body energies with respect to the 
$\Lambda\Lambda$ threshold and all three-body energies with respect to the 
$\Lambda\Lambda\alpha$ threshold.  That is, the energy available to the 
$\Xi N$ two-body and $\Xi N\alpha$ three-body channels is $E - \delta_{m}$, 
where $ \delta_{m} = m_{\Xi} + m_{N} - 2 m_{\Lambda}$; we understand $E$ 
to be the appropriate total two-body or three-body energy.

\section{Results and Discussion}\label{sec4}

Solution of the equations for the ground state of 
$^{\;\;\, 6}_{\Lambda\Lambda}$He was obtained by converting the above 
integrals 
equations into a set of coupled linear algebraic equations by replacing the 
integral with a Gauss-Legendre quadrature rule.  For the bound-state problem,
the kernel is smooth and well behaved, with no singularities on the real
momentum axis.  By setting $\tau_{\alpha}^{\Lambda \Xi}$ = 0, we can 
decouple the $\Lambda\Lambda\alpha$ and $\Xi N\alpha$ systems.  
In turning off the coupling in this way, the $\Lambda\Lambda$ interaction
included in the calculation of the binding energy of 
$^{\;\;\, 6}_{\Lambda\Lambda}$He is the $\Lambda\Lambda$ diagonal component 
of the $\Lambda\Lambda$--$\Xi N$ $T$-matrix from the solution of the 
coupled-channel problem. When coupling is included, we also include both 
$s$- and  $p$-wave $N\alpha$ interactions since both the
$P_{\frac{1}{2}}$ and $P_{\frac{3}{2}}$ channels support $N\alpha$ 
resonances.  All other potentials ({\it i.e.}, $\Lambda\Lambda$-$\Xi N$, 
$\Lambda\alpha$ and $\Xi\alpha$) are treated in s-wave only,
as this is expected to be the dominant component of these interactions. 
In Table~\ref{table10} we list all three-body channels for the  
$J^{\pi} = 0^{+}$, $T$ = 0 configuration.

We have argued above that only the $\Lambda\Lambda$ component of the full 
$\Lambda\Lambda$--$\Xi N$\ force is responsible for the binding of the 
hypernucleus, $^{\;\;\, 6}_{\Lambda\Lambda}$He.  That is, we suggest the 
$\Lambda\Lambda \rightarrow \Xi N$ transition is Pauli blocked and, as a 
consequence, only the $\Lambda\Lambda\alpha$ part of the Hilbert space 
is necessary to bind the hypernucleus.  In Table~\ref{table11} we summarize
results for $B_{\Lambda\Lambda}$ for each of the four potentials SA, SB, 
SC1 and SC2.  In each case we consider two different possibilities for 
the $\Lambda\alpha$\ interaction (SYNG and SIsle).  Although the 
$\Lambda\alpha$ interaction is uncertain, (the binding energy of 
$ ^{5}_{\Lambda}$He being the only experimental data to constrain the 
interaction) the  sensitivity of $B_{\Lambda\Lambda}$ to the type of 
$\Lambda\alpha$ potential used is minor and does not prevent our being 
able to discriminate among the types of $\Lambda\Lambda$--$\Xi N$ 
interaction considered.  As the results in Table~\ref{table11} demonstrate, 
the potentials that support two-body bound states ({\it i.e.}, SC1 and SC2), 
overbind $^{\;\;\, 6}_{\Lambda\Lambda}$He, even when only the 
$\Lambda\Lambda$ component of the force is included.  Potential SC1, in 
which the $\Lambda\Lambda$ system is bound by only $\sim$~0.7~MeV, 
predicts a value for $B_{\Lambda\Lambda}$ that is well above the 
experimental uncertainty associated with 
$B_{\Lambda\Lambda} = 10.9\pm 0.6$ MeV.  As we increase the strength of 
the $\Lambda\Lambda$--$\Xi N$ interaction to that of model SC2 (having a 
bound state of $\sim$~4.7 MeV), the value of $B_{\Lambda\Lambda}$ 
increases as expected, severely over binding the hypernucleus.  Based on 
these observations, we conclude that, if the experimental value of 
$B_{\Lambda \Lambda}$ is to be believed, then the $\Lambda\Lambda$--$\Xi N$ 
system cannot support a bound state (of at least 0.7 MeV) and still be 
consistent with the $\Lambda\Lambda$ separation energy in 
$^{\;\;\, 6}_{\Lambda\Lambda}$He. Finally, we note that the results for 
potential SB come closest to the experimental value.

In order to test our hypothesis that the 
$\Lambda\Lambda \leftrightarrow \Xi N$ conversion is suppressed in 
$^{\;\;\, 6}_{\Lambda\Lambda}$He, we performed bound-state calculations 
including this coupling term explicitly ({\it i.e.}, 
$\tau^{\Lambda\Xi}_\alpha\neq 0$). As we have previously suggested, 
we believe this conversion to be Pauli blocked, because the nucleon 
involved in the conversion cannot occupy the $1s$ state in a simple 
shell model picture.  For conversion to take place, either a nucleon 
in the core nucleus must be promoted into a positive parity excited state,
or the converted nucleon must go into the $2s$ state.  Both configurations
require $2\hbar\omega$ in energy and, as a result, are suppressed. To 
model this many-body effect within the two-body context of an 
$\alpha$-particle and a nucleon, we have parameterized the 
$S_{\frac{1}{2}}$  $N\alpha$\ interaction as repulsive.  Along with the 
$N\alpha$ $S_{\frac{1}{2}}$ interaction, we also include the attractive 
$N\alpha$ interactions in both the $P_{\frac{1}{2}}$ and $P_{\frac{3}{2}}$ 
channels, because both two-body interactions support resonances, and are 
coupled to the channel with the $N\alpha$ $S_{\frac{1}{2}}$ interaction. 
The results of these calculations are collected in Table~\ref{table12}, 
for various $\Lambda\Lambda$--$\Xi N$ and $\Lambda\alpha$ interactions. 
For the $N\alpha$ interactions we use model A of Ref.~\cite{esk92}, while 
for the $\Xi\alpha$ potential we utilize the interaction which we 
constructed for the $\Xi$-nucleus system (where the nucleus is $^4$He).

For each of the $\Lambda\Lambda$--$\Xi N$ interactions SA, SB, SC1 and 
SC2 there is a small increase in the binding energy 
$B_{\Lambda \Lambda}$, compared with those values listed in 
Table~\ref{table11}, and essentially independent of the type of 
$\Lambda\alpha$ interaction considered.   This is not surprising,
given that we have treated the $\alpha$ particle as an elementary 
object.   The incorporation of the ``Pauli principle'' for the 
$N\alpha$ $S_{\frac{1}{2}}$ interaction is not as obvious as it is 
in the shell-model.  However, the increase in $B_{\Lambda \Lambda}$ 
(0.23, 0.66, 1.38, and 2.36 MeV for potentials SA, SB, SC1, and SC2) is 
not as large as one might expect to see if the 
$\Lambda\Lambda \leftrightarrow \Xi N$ coupling was strong in the 
$^{\;\;\, 6}_{\Lambda\Lambda}$He hypernucleus.   It is evident from 
comparing results of Tables~\ref{table11} and \ref{table12} that
indeed the coupling is strongly suppressed, in contrast to that in
the two-body interactions in free space.  The inclusion of a repulsive 
$N\alpha$ $S_{\frac{1}{2}}$ interaction acts to prevent the $\Xi N\alpha$ 
part of the Hilbert space from adding much attraction to the three-body 
system.  Furthermore, we must conclude that the coupling between the 
three-body channels, $\Xi - ( N \alpha)_{S_{\frac{1}{2}}}$ in an 
$\cal{L}$ = 0 state, and 
$\Xi - ( N \alpha)_{P_{\frac{1}{2}},P_{\frac{3}{2}}}$ in
$\cal{L}$ = 1 states, is small; this is a result of the three-body dynamics. 
Because the p-wave $N\alpha$ interactions are strong, we would
expect them to provide significant additional attraction if these 
three-body channels were strongly coupled.

To emphasize our point, we have constructed effective single-channel 
$\Lambda\Lambda$ interactions designed to reproduce the scattering length 
$a_{\Lambda\Lambda}$ and effective range $r_{\Lambda\Lambda}$ for all
four coupled-channel interactions.  The parameters for these potentials 
SCSA, SCSB, SCSC1 and SCSC2, are listed in Table~\ref{table13}.
Comparing the values of $C_{\Lambda\Lambda}$ with those of the coupled
channel potential given in Table~\ref{table4} already gives an indication
that the effective potentials are considerably stronger in the 
$\Lambda\Lambda$ channel. The binding energy of 
$^{\;\;\, 6}_{\Lambda\Lambda}$He for the four effective potentials is 
given in Table~\ref{table14}, and are all much larger than those shown 
in Table~\ref{table11}, except for the weakly attractive potential SA.
For potential SCSB the increase in binding due to this effective 
$\Lambda\Lambda$ interaction (compared with potential SB) is $\sim$ 
2.5 MeV.  This is to be compared with an increase of $\sim 0.7$~MeV 
when including the $\Lambda\Lambda$--$\Xi N$ coupling. This difference 
reflects the extra binding a free space $\Lambda\Lambda$ interaction 
would give for $B_{\Lambda\Lambda}$.  

Others have included repulsive dispersive three-body forces, along with
two-body forces, to eliminate such overbinding in $\Lambda$ hypernuclei.  
In particular Bodmer {\it et al}.~\cite{bod84,bod88,bod87,usm95} have 
included dispersive $\Lambda NN$ three-body forces.  They found it 
necessary to include repulsive three-body forces to avoid overbinding 
for $^5_{\Lambda}$He.  However, if the effects of the suppression of 
the $\Lambda N$--$\Sigma N$ coupling are included explicitly, then 
one can account for the experimental value of $B_{\Lambda}(^5_{\Lambda}$He) 
using a reasonable model of the $\Lambda N$-$\Sigma N$ interaction 
\cite{gib94}, without the need to introduce significant three-body force 
effects.  Indeed, we find for $^{\;\;\, 6}_{\Lambda\Lambda}$He 
that we can account for the binding energy within a realistic model of the 
$\Lambda\Lambda$--$\Xi N$ interaction (one comparable in strength to that 
of the $nn$ interaction) without the need to include $\Lambda\Lambda N$ 
three-body forces.  That is, we have shown that incorporation of important 
two-body effects (inclusion of $\Lambda\Lambda$--$\Xi N$ coupling in the 
two-body, and the resultant Pauli blocking in the $A=6$ hypernucleus) 
can reduce the value of $B_{\Lambda\Lambda}$ to lie within acceptable 
limits of the available experimental datum.

\section{Summary and Conclusions}

Based upon the information contained in Tables~\ref{table11} and
\ref{table12} in particular, we conclude that it is primarily the 
$\Lambda\Lambda$ component of the $\Lambda\Lambda$--$\Xi N$ force 
that is sampled in the $^{\;\;\, 6}_{\Lambda\Lambda}$He hypernucleus.  
Furthermore, we observe that potentials SC1 and SC2 are unlikely 
to be realistic models for the interaction if we are to believe 
the experimental value for $B_{\Lambda \Lambda}$.  
Potential SA, on the other hand, 
predicts a value for $B_{\Lambda\Lambda}$ that lies $\sim$ 0.8 MeV 
below the lower limit for the experimental value (for the SYNG model 
of the $\alpha\Lambda$ interaction).  Therefore, it cannot be completely
discounted.  However, potential SB (the model that gives rise to an 
anti-bound state similar to that in the $nn$ system), leads to a value 
for $B_{\Lambda\Lambda}$ that is $\sim$ 0.1 MeV above the upper limit 
of the experimental value, and is, hence, in reasonable agreement with 
experiment.  This potential yields
\begin{equation}
  a_{\Lambda \Lambda} \simeq -21.0 \hspace{2 mm} \mbox{fm}\ ,\label{eq:36}
\end{equation}
which is comparable to the $ ^{1}S_{0}$ $nn$ scattering length.  
That is, we find agreement with experiment for a model in which the
$\Lambda\Lambda$--$\Xi N$ interaction and the $nn$ interaction 
have comparable overall strength.  We emphasize that it is the full 
$\Lambda\Lambda$--$\Xi N$ interaction that gives rise to a value of 
$a_{\Lambda \Lambda}$ $\simeq$ -21.0 fm.   Even though in free space
the SB model predicts such a value for $a_{\Lambda \Lambda}$, it can
still give a value of
\begin{equation}
  - \langle V_{\Lambda \Lambda} \rangle \simeq 4.7 
                                \hspace{1 mm} \mbox{MeV} \label{eq:37}
\end{equation}
in good agreement with the $^{\;\;\, 6}_{\Lambda\Lambda}$He datum.  This 
value for $ - \langle V_{\Lambda \Lambda} \rangle$ is significantly smaller
than a similar $nn$ interaction quantity of
\begin{equation}
- \langle V_{n n} \rangle \simeq 6 - 7 \hspace{1 mm} 
                      \mbox{MeV} \ .                     \label{eq:38}
\end{equation}
It is the suppression of the $\Lambda\Lambda \leftrightarrow \Xi N$ 
coupling that is responsible for the reduction in 
$ - \langle V_{\Lambda \Lambda} \rangle$ compared with 
$ - \langle V_{n n} \rangle$.  In other words, there can exist a
$\Lambda\Lambda$--$\Xi N$ interaction that is consistent with the
experimental data for $B_{\Lambda \Lambda}$, and at the same time allows
for a free space value for $a_{\Lambda \Lambda}$ which is comparable to
$a_{nn}$. 

\acknowledgments
The authors would like to dedicate the present work to the memory of
Carl Dover who played a central role in initiating this investigation
and in particular the construction of the OBE potential in the 
strangeness -2 sector.
The work of SBC and IRA was supported by the Australian Research
Council.  That of BFG was performed under the auspices of the U.~S. 
Department of Energy.

\newpage
\appendix 
\section{The OBE potential}

In this Appendix we outline the one boson exchange potential in the
$S=-2$ sector. Since we have restricted the analysis to $s$-wave only, 
the potential has the form given in Eq.~(\ref{eq:6}) with the central 
and spin-dependent components of the form
\begin{eqnarray}
V_\xi &=& V_\xi^{\rm S} + V_\xi^{\rm PS} + V_\xi^{\rm V}\qquad \xi=c,\sigma
                                                \nonumber \\
      &=& \sum_i\ V_\xi^{(i)}\ \phi(x_i)\ ,      \label{eq:A1}
\end{eqnarray}
where the $i$ sum runs over all possible exchanged mesons. The Nijmegen
potential D~\cite{nijD} includes the complete nonets of pseudoscalar (PS)
$[\pi,
\eta, \eta', K]$ and vector (V) $[\rho, \omega, \phi, K^*]$ mesons as well as
the exchange of  the unitary singlet scalar $\epsilon$ meson. The function
$\phi(x_i)$ is the standard Yukawa function including the soft cutoff, {\it
i.e.}
\begin{equation}
\phi(x_i) = \left[\frac{e^{-x_i}}{x_i} - {\cal C}\ \left(\frac{\cal M}{m_i}
\right)\ \frac{e^{{-\cal M}r}}{{\cal M}r}\right]\ , \label{eq:A2}
\end{equation}
where $x_i = m_i r$. The meson masses $m_i$ are given in Table~\ref{table15}
and \ref{table16}, while the cutoff mass for all meson exchanges is taken 
to be ${\cal M}=2500$~MeV. 
The cutoff strengths ${\cal C}$ for the different potentials are given in 
Table~\ref{table1}. For the present potential we follow Dover and 
Gal~\cite{dov83} and introduce 
average $K$ and $K^*$ masses given by $\bar{m}_K^2 = m_K^2- (1/2)
[(m_\Xi-m_\Lambda)^2 + (m_N-m_\Lambda)^2]$, with a similar expression for
$\bar{m}_{K^*}^2$. To give the $\rho$ and $\epsilon$
mesons width, the Nijmegen group~\cite{nijD} have introduced two $\rho$ and 
two $\epsilon$ masses and have taken the linear combination of the exchange of
the two mesons; {\it e.g.} for the $\epsilon$ we utilize
\begin{equation}
V_\xi^{(\epsilon)}\,\phi(m_\epsilon r)\rightarrow 
         \beta_1\,V_\xi^{\epsilon_1}\,\phi(m_{\epsilon_1}r) + 
         \beta_2\,V_\xi^{\epsilon_2}\,\phi(m_{\epsilon_2}r)   \label{eq:A3}
\end{equation}
and make a similar substitution for $\rho$ meson exchange. The values 
of the $\beta_i$ and $m_i$ are given in Table~\ref{table15}. To guarantee 
that the cutoff is always repulsive, we take ${\cal C}<0$ if $V^{(i)}_\xi >0$.

The one boson exchange potential, presented diagrammatically in
Figs.\ 1 and 2, for the reaction
\begin{equation}
B_1 + B_2 \rightarrow B_3 + B_4 \ .                        \label{eq:A4}
\end{equation}
takes the form given in Eq.~(\ref{eq:A1}), with the strength $V_c^{(i)}$ 
for the central part of the potential given by\cite{dovpr}:
\begin{eqnarray}
  V^{(i)}_{c}&=& 0 \hspace{7.6cm} \mbox{for PS exchange}\ ,  \label{eq:A5}\\
  V^{(i)}_{c}&=& - \frac{g_{B_{1}B_{3}}^{(i)}\  g_{B_{2}B_{4}}^{(i)}}{4\pi}
       \ m_{i}\ \left(\ 1-\frac{m_{i}^{2}}{8 M_{1}M_{2}}\ \right)
       \qquad\qquad \mbox{for S exchange}    \ ,            \label{eq:A6}\\ 
  V^{(i)}_{c} &=& \frac{g_{B_{1}B_{3}}^{(i)}\ g_{B_{2}B_{4}}^{(i)}}{4\pi} 
       \ m_{i}\ \left[\ 1 + \left( \frac{f}{g}\right)_{B_{1}B_{3}}^{(i)}  
       \ \frac{m_{i}^{2}}{4 M M_{1}} 
       + \left( \frac{f}{g} \right)_{B_{2}B_{4}}^{(i)}\ 
       \frac{m_{i}^{2}}{4 M M_{2}} \right. \nonumber\\ 
       &&\qquad\left.  + \ \left( \frac{f}{g} \right)_{B_{1}B_{3}}^{(i)}\  
       \left(\frac{f}{g} \right)_{B_{2}B_{4}}^{(i)}
       \frac{m_{i}^{4}}{16 M^{2} M_{1} M_{2}} \right]
                      \quad \mbox{for V exchange} \ ,       \label{eq:A7}
\end{eqnarray}
for the exchanged mesons. The spin-dependent part of the potential, 
$V_{\sigma}^{(i)}$, is given by\cite{dovpr}: 
\begin{eqnarray}
  V^{(i)}_{\sigma}&=& \frac{g_{B_{1}B_{3}}^{(i)}\ g_{B_{2}B_{4}}^{(i)}}{4\pi}  
        \frac{m_{i}^{3}}{12 M_{1} M_{2}} 
        \hspace{6cm}\mbox{for PS exchange}                    \label{eq:A8}\\
  V^{(i)}_{\sigma}&=& 0 \hspace{9.5cm} \mbox{for S exchange}  \label{eq:A9}\\
  V^{(i)}_{\sigma}&=& \frac{g_{B_{1}B_{3}}^{(i)}\ g_{B_{2}B_{4}}^{(i)}}{4\pi} 
         \ \frac{m_{i}^{3}}{6 M_{1} M_{2}}\
         \left[\ 1 + \left( \frac{f}{g} \right)_{B_{1}B_{3}}^{(i)}\frac{M_{1}}{M}
       + \left( \frac{f}{g} \right)_{B_{2}B_{4}}^{(i)}
         \frac{M_{2}}{M} \right.                            \nonumber \\
       && \left. + \left( \frac{f}{g} \right)_{B_{1}B_{3}}^{(i)}\ 
         \left( \frac{f}{g} \right)_{B_{2}B_{4}}^{(i)}
         \frac{M_{1} M_{2}}{M^{2}}\ \left(\ 1+\frac{m_{i}^{2}}{8 M_{1}
          M_{2}}  \ \right) \ \right] \
         \qquad\mbox{for V exchange}\ .                     \label{eq:A10} 
\end{eqnarray} 
In Eqs.~(\ref{eq:A9}) and (\ref{eq:A10}) the coupling constants 
$g_{B_{1}B_{3}}^{(i)}$ and $f_{B_{1}B_{3}}^{(i)}$ are the electric and
magnetic couplings of meson $i$ to baryons 1 and 3, $m_{i}$ is the mass 
of the exchanged meson, $M_{1} = (M_{B_{1}}M_{B_{3}})^{\frac{1}{2}}$ and 
$M_{2} = (M_{B_{2}}M_{B_{4}})^{\frac{1}{2}}$.  Here, M is a scale mass, 
assumed to be the mass of the nucleon ($M_{N}$). The coupling constants 
that we need for these potentials can be found in Table \ref{table16}.

For strange meson exchange, the potential has the same basic form as
$V_{\mbox{\scriptsize OBE}}(r)$ ({\it i.e} Eq.~(\ref{eq:6})), with the 
requirement that we replace
\mbox{\boldmath $t_{1} \cdot t_{2}$} with the isospin factor
$\sqrt{2}$.  That is, formally we have \cite{dov83} 
\begin{equation}
  \langle \Lambda\Lambda |\,{\bf t}_{1} \cdot {\bf t}_{2} 
  | \Xi~N \rangle_{t=0} = \sqrt{2} \ , \label{eq:A.11}
\end{equation} 
and introduce space and spin exchange operators, $P_{x}$ and $P_{\sigma}$
respectively. For the $ ^{1}S_{0}$ state we have $P_{x}$ = 1 and
$P_{\sigma}$ = -1.

\begin{figure}
\vskip 1 cm

\centerline{\epsfig{figure=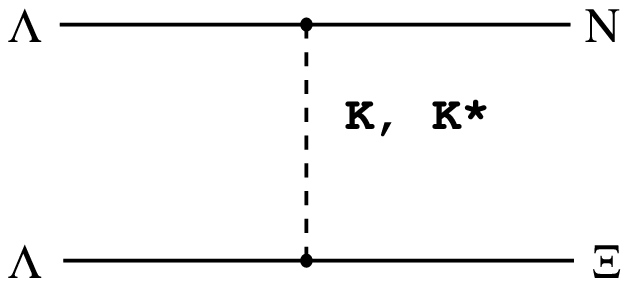,width=8cm}}  

\noindent Fig. 1.  The first order meson exchange diagram for the
$\Lambda\Lambda \rightarrow \Xi N$ process.  \label{fig1}
\vskip 2.0 cm

\centerline{\epsfig{figure=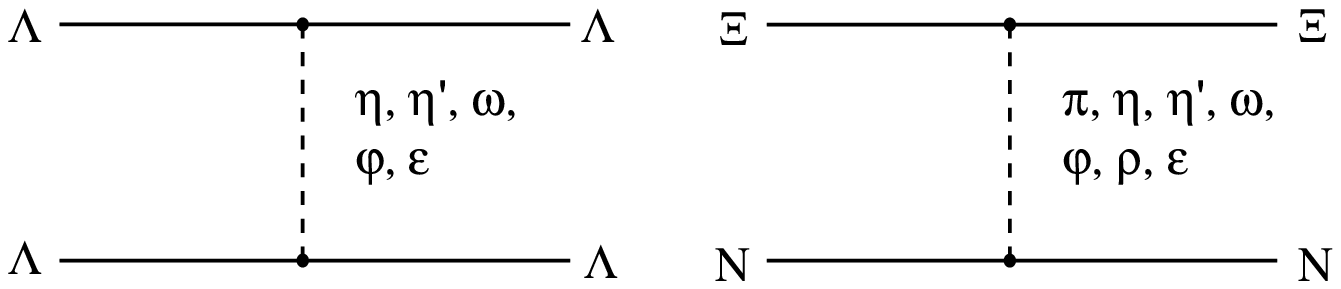,width=15cm}}
\noindent Fig. 2. The first order meson exchange diagrams for the
$\Lambda\Lambda$ and $\Xi N$ systems.\label{fig2} 
\end{figure}

\newpage

\begin{table}
\begin{minipage}{8.5cm}
\centering
\caption{The cutoff parameter ${\cal C}$ for the four $\Lambda\Lambda$--$\Xi N$
  coupled-channels potentials and the $\Xi N$ single-channel 
          potentials specified.     \label{table1}}   
\vspace{12pt}
\begin{tabular}{ccc|cc}
  Pot. & ${\cal C}$  & &    Channel       & ${\cal C}$  \\ \hline
    A  &  157.0      & & $^{1}S_{0}$ t=1  &  28.65      \\
    B  &   87.0      & & $^{3}S_{1}$ t=0  &  6.70       \\
    C1 &   66.0      & & $^{3}S_{1}$ t=1  &  10.9       \\ 
    C2 &   50.2      & & &  \\ 
\end{tabular}
\end{minipage}
\end{table}

\begin{table}
\centering
\caption{{\protect
     The effective range parameters for the $\Lambda\Lambda$--$\Xi N$ 
     coupled-channels potentials in the $^{1}S_{0}$ partial wave.} 
     \label{table2}}    
\vspace{12pt}
\begin{tabular}{cccccc} 
  Pot. &  $a_{\Lambda\Lambda}$~(fm) & $r_{\Lambda\Lambda}$~(fm) &  
  $a_{\Xi N}$~(fm) & $r_{\Xi N}$~(fm) &  B.E. (MeV)  \\ \hline
   A  &  -1.91 & 3.36 &  -2.12-0.75i & 3.45-0.45i  & UB   \\
   B  &  -21.1 & 1.86 &  -2.05-6.53i & 2.12-0.21i  & UB   \\
   C1 &   7.82 & 1.41 &   3.08-5.26i & 1.74-0.144i & 0.71 \\
   C2 &   3.37 & 1.0  &   3.37-2.54i & 1.44-0.10i  & 4.74 \\ 
\end{tabular}
\end{table}

\begin{table}
\begin{minipage}{8.5cm}
\centering
\caption{\noindent The effective range parameters for the $\Xi N$ 
potentials in the various channels. \label{table3}} 
\vspace{12pt}
\begin{tabular}{ccc} 
    channel        &  $a_{\Xi N}$~(fm) & $r_{\Xi N}$~(fm)  \\ \hline
   $^{1}S_{0}$ t=1 &  -1.94       & 3.00         \\
   $^{3}S_{1}$ t=0 &  -1.75       & 3.20         \\
   $^{3}S_{1}$ t=1 &  -1.81       & 3.28         \\ 
\end{tabular}
\end{minipage} 
\end{table}

\begin{table}
\centering
\caption{ The parameters of the $^{1}S_{0}$ $\Lambda\Lambda$--$\Xi N$ 
 coupled-channel separable potentials. The $\beta$s are in fm$^{-1}$
while the $C_{\alpha\alpha'}$ are in fm$^{-2}$. \label{table4}} 
\vspace{12pt}
\begin{tabular}{lccccc} 
   Pot. & $C_{\Lambda\Lambda}$ & $\beta_{\Lambda}$ & 
$C_{\Xi \Xi}$ & $\beta_{\Xi}$ & $C_{\Lambda \Xi}$  \\ \hline
   SA & -0.25874 & 1.3202 & -0.32119 & 1.3301 & 0.1820  \\
   SB & -0.28005  & 1.1719 & -0.44461  & 1.2749 & 0.22015  \\
   SC1 & -0.88154 & 1.5834 & -2.1590 & 2.0433 & 0.57062  \\
   SC2 & -1.4568 & 1.7752 & -0.85766 & 1.3789 & 0.55982  \\ 
\end{tabular}
\end{table}

\begin{table}
\begin{minipage}{8.5cm}
\centering
\caption{ The parameters of the $\Xi N$ separable potentials for 
the given channels. The units are the same as in Table~\ref{table4}.
\label{table5} } 
\vspace{12pt}
\begin{tabular}{ccc} 
   channel   & $C_{\Xi N}$ & $\beta_{\Xi}$ \\ \hline
  $^{1}S_{0}$ t=1  & -0.42588 & 1.4681  \\
  $^{3}S_{1}$ t=0  & -0.37703 & 1.4352  \\
  $^{3}S_{1}$ t=1  & -0.34869 & 1.3969  \\ 
\end{tabular}
\end{minipage}
\end{table}

\begin{table}
\centering
\caption{{The effective range parameters for the $\Lambda\Lambda$--$\Xi N$ 
 coupled-channels separable potentials in the $^{1}S_{0}$ partial wave.}
\label{table6} }    
\vspace{12pt}
\begin{tabular}{lccccc} 
  Pot. & $a_{\Lambda\Lambda}$~(fm) & $r_{\Lambda\Lambda}$~(fm) & 
       $a_{\Xi N}$~(fm) &  $r_{\Xi N}$~(fm) & B.E. (MeV) \\ \hline
    
    SA   & -1.90 & 3.33 & -2.08-0.81i & 3.44-0.22i & UB \\
      
    SB   & -21.0 & 2.54 & -2.07-6.52i  & 2.62-0.15i & UB \\

    SC1   & 7.84 & 1.48 & 3.05-5.28i & 1.45+0.074i & 0.71 \\
        
    SC2   & 3.36 & 1.0 & 3.35-2.50i  & 1.83-0.10i & 4.73 \\ 
\end{tabular}
\end{table}


\begin{table}
\begin{minipage}{8.5cm}
\centering
\caption{ The parameters of the $\Lambda N$ central YNG interaction.
The strengths are in MeV.\label{table7} }   
\vspace{12pt}
\begin{tabular}{cc|ccc} 
& Range $\beta_i$ (fm$^{-1}$) & 1.5 & 0.9 & 0.5 \\ \hline
 &  a & -13.21 & -555.2 & 2686.0 \\
$ ^{1}S_{0}$ &  b & 5.705 & 408.1 & -2137.0  \\
 &  c & -2.412  & -135.5 & 813.6 \\ \hline 
 &  a & -13.57 & -389.7 & 1670.0 \\
$ ^{3}S_{1}$ & b & 5.873 & 295.7 & -1337.0  \\
 &  c & -1.989  & -99.82 & 519.7 \\ 
\end{tabular}
\end{minipage}
\end{table}

\begin{table}
\begin{minipage}{8.5cm} 
\centering
\caption{{The effective range parameters and the binding energy for the
  $S_{\frac{1}{2}}$ $\Lambda\alpha$ and  $S_{\frac{1}{2}}$ $\Xi\alpha$
  interactions.} \label{table8} }   
\vspace{12pt}
\begin{tabular}{c|ccc|c} 
 Pot. &  $a$~(fm) & $r$~(fm)  & & B.E.~(MeV) \\ \hline
 YNG  & 4.27 & 2.11 & & 3.1  \\
 Isle & 4.23 & 2.10 & & 3.1  \\ \hline
 DG   & 4.53 & 1.98 & & 2.085 \\
\end{tabular}
\end{minipage}
\end{table}

\begin{table}
\begin{minipage}{8.5cm}
\centering
\caption{ The parameters of the $N\alpha$ separable potentials for the 
various two-body channels. The $\beta$s are in fm$^{-1}$ while the 
$C^\ell_{N\alpha}$ are in fm$^{-2}$ for the $s$-wave and fm$^{-4}$ 
for the $p$-waves.\label{table9} } 
\vspace{12pt}
\begin{tabular}{c|cc}
  Channel & $C_{N \alpha}^{\ell}$ & $\beta_{N \alpha}^{\ell}$ \\ \hline
  $S_{\frac{1}{2}}$ & 1.0519 & 0.7496 \\
  $P_{\frac{1}{2}}$ & -1.8221 & 1.1770 \\
  $P_{\frac{3}{2}}$ & -7.9735 & 1.4490 \\ 
\end{tabular}
\end{minipage}
\end{table}

\begin{table}
\begin{minipage}{8.5cm}
\centering
\caption{ Three-body channels allowed for the $J^{\pi}$ = $0^{+}$, $T$ = 0 
configuration. \label{table10} }
\vspace{12pt}
\begin{tabular}{c|c|cccccc} 
& Channel & t & s & $\ell$ & j & $\cal{S}$ & $\cal{L}$ \\ \cline{1-8}
$\underline{\Lambda \Lambda \alpha}$: & & & & & & & \\
& $\Lambda\alpha (S_{\frac{1}{2}}$) & 0 & $\frac{1}{2}$ & 0 & $\frac{1}{2}$ 
& 0 & 0 \\
& $\Lambda\Lambda (^{1}S_{0}$) & 0 & 0 & 0 & 0 & 0 & 0 \\ \cline{1-8}
$\underline{\Xi N \alpha}$: & & & & & & & \\ 
& $\Xi N (^{1}S_{0}$) & 0 & 0 & 0 & 0 & 0 & 0 \\
& $N\alpha (S_{\frac{1}{2}}$) & $\frac{1}{2}$ & $\frac{1}{2}$ & 0 & 
$\frac{1}{2}$ & 0 & 0 \\ 
& $N\alpha (P_{\frac{1}{2}}$) & $\frac{1}{2}$ & $\frac{1}{2}$ & 1 & 
$\frac{1}{2}$ & 1 & 1 \\ 
& $N\alpha (P_{\frac{3}{2}}$) & $\frac{1}{2}$ & $\frac{1}{2}$ & 1 & 
$\frac{3}{2}$ & 1 & 1 \\ 
& $\Xi\alpha (S_{\frac{1}{2}}$) & $\frac{1}{2}$ & $\frac{1}{2}$ & 0 & 
$\frac{1}{2}$ & 0 & 0 \\  
\end{tabular}
\end{minipage}
\end{table}

\begin{table}
\begin{minipage}{8.5cm}
\centering
\caption{The $\Lambda \Lambda$ separation energy with only the 
  $\Lambda\Lambda$ component of the full interaction included. 
  \label{table11}}    
\vspace{12pt}
\begin{tabular}{c|ccc|c}
Channel & $\Lambda \Lambda$ & $\Lambda \alpha S_{\frac{1}{2}}$ & &
$B_{\Lambda\Lambda}$ (MeV) \\ \hline
model & SA  & SYNG  & & 9.5078 \\
      & SA  & SIsle & & 9.8110 \\ \cline{2-5}
      & SB  & SYNG  & & 11.606 \\
      & SB  & SIsle & & 11.929 \\ \cline{2-5}
      & SC1 & SYNG  & & 14.533 \\ 
      & SC1 & SIsle & & 14.890 \\ \cline{2-5}
      & SC2 & SYNG  & & 17.508 \\
      & SC2 & SIsle & & 17.889 \\
\end{tabular}
\end{minipage}
\end{table}

\begin{table}
\centering
\caption{The $\Lambda\Lambda$ separation energy for the full 
  $\Lambda\Lambda$-$\Xi N$ interaction. \label{table12}}  
\begin{tabular}{c|ccccccc|c}
Channel  & $\Lambda\Lambda$-$\Xi N$ & $\Lambda\alpha S_{\frac{1}{2}}$ & 
$N \alpha S_{\frac{1}{2}}$ & $N \alpha P_{\frac{1}{2}}$ & 
$N \alpha P_{\frac{3}{2}}$ & $\Xi\alpha S_{\frac{1}{2}}$ & &
$B_{\Lambda\Lambda}$ (MeV) \\ \hline
model  & SA & SYNG   & A & A & A & SDG & & 9.7381 \\
       & SA & SIsle  & A & A & A & SDG & & 10.043 \\ \cline{2-9}
       & SB & SYNG   & A & A & A & SDG & & 12.268 \\
       & SB & SIsle  & A & A & A & SDG & & 12.594\\  \cline{2-9}
       & SC1 & SYNG  & A & A & A & SDG & & 15.912 \\
       & SC1 & SIsle & A & A & A & SDG & & 16.268 \\ \cline{2-9}
       & SC2 & SYNG  & A & A & A & SDG & & 19.836 \\
       & SC2 & SIsle & A & A & A & SDG & & 20.507 \\
\end{tabular}
\end{table}

\begin{table}
\begin{minipage}{8.5cm}
\centering
\caption{ The parameters of the $^{1}S_{0}$ $\Lambda \Lambda$ single channel
potentials. The units of $\beta_\Lambda$ and $C_{\Lambda\Lambda}$ are
the same as in Table~\ref{table4}. \label{table13} }  
\vspace{12pt}
\begin{tabular}{c|cc}
  Pot. & $C_{\Lambda\Lambda}$ & $\beta_{\Lambda}$ \\ \hline
   SCSA    & -0.3148  & 1.3534 \\
   SCSB    & -0.9995  & 1.6738  \\
   SCSC1   & -1.8960  & 1.9407  \\
   SCSC2   & -4.7699  & 2.5310  \\
\end{tabular}
\end{minipage}
\end{table}

\begin{table}
\begin{minipage}{8.5cm}
\centering
\caption{The $\Lambda\Lambda$ separation energy for various single-channel 
$\Lambda\Lambda$ interactions. \label{table14}} 
\vspace{12pt}
\begin{tabular}{c|ccc|c}
Channel & $\Lambda \Lambda$ & $\Lambda \alpha S_{\frac{1}{2}}$ & &
$B_{\Lambda\Lambda}$ (MeV) \\ \hline
Model 
 & SCSA     & SYNG  & & 10.007 \\
 & SCSA     & SIsle & & 10.317 \\ \cline{2-5}
 & SCSB     & SYNG  & & 14.138 \\
 & SCSB     & SIsle & & 14.494 \\ \cline{2-5}
 & SCSC1    & SYNG  & & 17.842 \\ 
 & SCSC1    & SIsle & & 18.225 \\ \cline{2-5}
 & SCSC2    & SYNG  & & 23.342 \\
 & SCSC2    & SIsle & & 23.750 \\
\end{tabular}
\end{minipage}
\end{table}


\begin{table}
\begin{minipage}{8.5cm}
\centering
\caption{The  weighting parameters $\beta_i$ and the meson masses $m_i$ 
in MeV used to simulate a width for the $\rho$ and $\epsilon$ mesons.
\label{table15}}
\vspace{12pt}
\begin{tabular}{lcccc}
Meson      & $\beta_1$  &   $m_1$   &  $\beta_2$  &  $m_2$  \\ \hline
$\rho$     & 0.15874    & 628.74    &  0.78321    & 878.18  \\
$\epsilon$ & 0.19986    & 508.52    &  0.55241    & 1043.79 \\
\end{tabular}
\end{minipage}
\end{table}

\begin{table}
\centering
\caption{ Coupling constants $g^{(m)}_{BB'}$ and $f^{(m)}_{BB'}$ for PS, V 
and S meson exchanges, for Nijmegen model D. Also included are the masses of 
the exchanged mesons in MeV.\label{table16}}  
\vspace{12pt}
\begin{tabular}{c|rrrrrr}
  Meson $m$ & Mass & $NN\,m$ & $\Xi\Xi\,m$ & $\Lambda\Lambda\,m$ &
  $\Lambda\Xi\,m$ & $\Lambda N\,m$\\ \hline
  $\pi$ \hspace{2.0 mm}          $g$&138.03& 3.66& -0.11 &       &       & \\
  $\eta$ \hspace{2.0 mm}         $g$&548.8 & 2.73& -3.289& -1.361&       & \\
  $\eta^{\prime}$ \hspace{2.0 mm}$g$&957.57& 3.89&  5.015&  4.639&       & \\
  $K$ \hspace{2.0 mm}            $g$&457.83&     &       &       & 1.986 & -4.16 \\ \hline
  $\rho$ \hspace{2.0 mm}         $g$&      &0.594& 0.594 &       &       & \\
                                 $f$&      &4.817& -1.60 &       &       &  \\
  $\phi$ \hspace{2.0 mm}         $g$&1019.5&-1.124&-2.806& -1.965&       & \\
                                 $f$&      &-0.51 & -5.06& -4.298&       & \\
  $\omega$ \hspace{2.0 mm}       $g$&782.6 & 3.373& 2.184& 2.779 &       & \\
                                 $f$&      & 2.34 &-0.878& -0.338&       & \\
  $K^{*}$ \hspace{0.1 mm}        $g$&871.63&      &      &       & 1.032 & -1.032 \\ 
                                 $f$&      &      &      &       & 0.934 & -4.64 \\ \hline
  $\epsilon$ \hspace{2.0 mm}     $g$&      & 5.032& 5.032& 5.032 &       & \\ 
\end{tabular}
\end{table}


\newpage
\begin{references}
\bibitem{dan63} M.~Danysz, K.~Garbowska, J.~Pniewski, T.~Pniewski,
                J.~Zakrewski, E.~R.~Fletcher, {\it et al.},
                Phys.~Rev.~Lett.~{\bf 11}, 29 (1963); 
                Nucl.~Phys.~{\bf 49}, 121 (1963).
\bibitem{pro66} D.~J.~Prowse, Phys.~Rev.~Lett.~{\bf 17}, 782 (1966).
\bibitem{aok91} S.~Aoki, S.~Y.~Bahk, K.~S.~Chung, S.~H.~Chung,
                H.~Funahashi, C.~Hahn, {\it et al.}\
                Prog.~Theo.~Phys.~{\bf 85}, 1287 (1991).
\bibitem{dov91} C.~B.~Dover, D.~J.~Millener, A.~Gal, and 
                D.~H.~Davis, Phys.~Rev.~C {\bf 44}, 1905 (1991).
\bibitem{wil59} D.\ H.\ Wilkinson, S.\ J.\ St.\ Lorant, D.\ K.\
                Robinson, and S.\ Lokanathan, Phys.\ Rev.\ Lett.\
                {\bf 3}, 397 (1959).
\bibitem{bar63} W.\ H.\ Barkas, M.\ A.\ Dyer, and H.\ H.\ Heckman, 
                Phys.\ Rev.\ Lett.\ {\bf 11}, 429 (1963).
\bibitem{bho63} B.\ Bhownik, Nuovo Cimento {\bf 29}, 1 (1963).
\bibitem{bec68} A.\ Bechdolff, G.\ Baumann, J.\ P.\ Gerber, and
                P.\ C\"uer, Phys.\ Lett.\ B {\bf 26}, 174 (1968).
\bibitem{cat69} J.\ Catala, F.\ Senet, A.\ F.\ Tejerina, and E.\ 
                Villar, in {\it Proceedings of the International
                Conference on Hypernuclear Physics}, Vol. 2, ed. by 
                A.\ R.\ Bodmer and L.\ G.\ Hyman (Argonne 1969) p.\
                758.
\bibitem{mon79} A.\ S.\ Mondal, A.\ K.\ Basak, M.\ M.\ Kasim, and 
                A.\ Husain, Nuovo Cimento A {\bf 54}, 333 (1979).
\bibitem{tor96} See W.\ Tornow, H.\ Wita$\l$a, and R.\ T.\ Braun, 
                Few-Body Systems {\bf 21}, 97 (1996) and the references
                cited therein.
\bibitem{dal89} R.~H.~Dalitz, D.~H.~Davis, P.~H.~Fowler, A.~Montwill,
                J.~Pniewski, and J.~A.~Zakrewski, Proc.~Roy.~Soc.\ 
                London {\bf A426}, 1 (1989).
\bibitem{swa71} J.\ J.\ de Swart, M.\ M.\ Nagels, T.\ A. Rijken, and
                P.\ A.\ Verhoeven, {\it Springer Tracts in Modern
                Physics} {\bf 60}, 138 (1971).
\bibitem{nijD}  M.\ M.\ Nagels, T.\ A.\ Rijken, and J.\ J.\ de
                Swart, Phys.\ Rev.\ D {\bf 15}, 2547 (1977).
\bibitem{nijF}  M.~M.~Nagels, T.\ A.\ Rijken, and J.\ J,\
                de Swart,  Phys.~Rev.~D {\bf 20}, 1633 (1979);
                M.\ M.\ Nagels, T.\ A. Rijken, and J.\ J.\ de Swart,
                Ann. Phys. (NY) {\bf 79}, 338 (1973).
\bibitem{nijSC} P.~M.~M.~Maessen, T.\ A.\ Rijken and J.\ J.\
                de Swart, Phys.\ Rev.\ C {\bf 40}, 2226 (1989).
\bibitem{swa94} J.\ J.\ de Swart, Proceedings of the {\it U.\ S.\ -- 
                Japan Seminar on the Hyperon-Nucleon Interaction}, ed.\
                by B.\ F.\ Gibson, P.\ D.\ Barnes, and K.\ Nakai,
                (World Scientific, Singapore, 1994), pp.\ 37-54.
\bibitem{tim94}  R.\ Timmermans, Proceedings of the {\it U.\ S.\ -- 
                Japan Seminar on the Hyperon-Nucleon Interaction}, ed.\
                by B.\ F.\ Gibson, P.\ D.\ Barnes, and K.\ Nakai,
                (World Scientific, Singapore, 1994), pp.\ 179-188.
\bibitem{jul89} B.~Holzenkamp, K.\ Holinde, and J.\ Speth, 
                Nucl.~Phys.~{\bf A500}, 485 (1989).
\bibitem{hol92} K.~Holinde,  Nuc.~Phys.~{\bf A547}, 255c (1992).
\bibitem{jul92} A.~G.~Reuber, K.\ Holinde, and J.\ Speth, Czech.\ 
                J.~Phys.~{\bf 42}, 1115 (1992).
\bibitem{reu94} A.\ G.\  Reuber, Proceedings of the {\it U.\ S.\ -- 
                Japan Seminar on the Hyperon-Nucleon Interaction}, ed.\
                by B.\ F.\ Gibson, P.\ D.\ Barnes, and K.\ Nakai,
                (World Scientific, Singapore, 1994), pp.\ 159-168.
\bibitem{jur73} M.\ Juric, G.\ Bohm, J.\ Klabuhn, U.\ Krecker,
                F.\ Wysotzki, G.\ Coremans-Bertand, J.\ Sacton, G.\ Wilquet,
                T.\ Cantrell, F. Esmael, A.\ Montwill, D.\ H.\ Davis,
                {\it et al.}\, Nucl.\ Phys.\ {\bf B52}, 1 (1973).
\bibitem{dav67} D.~H.~Davis and J.~Sacton, in {\it High Energy Physics}, 
                Vol. II, (Academic Press, New York, 1967) p.~365.
\bibitem{dav91} D.\ H.\ Davis, 	Proceedings of the {\it LAMPF Workshop on
                ($\pi,K$) Physics}, A.\ I.\ P.\ Conf.\ Proc.\ {\bf 224},
                ed.\ by B.\ F.\ Gibson, W.\ R.\ Gibbs, and M.\ B.\ Johnson,
               (American Institute of Physics, New York, 1991), pp.\ 38-48.
\bibitem{bam73} A.\ Bamberger, M.\ A.\ Faessler, U. Lynn, H. Piekarz,
                J.\ Piekarz, J.\ Pniewski, B.\ Povh, H.\ G.\ Ritter, and V.\
                Soergel, Nucl.\ Phys.\ {\bf B60}, 1 (1973).
\bibitem{bej79} M.\ Bejidian, E.\ Descroix, J.\ Grossiord, A.\
                Guichard, M.\ Gusakow, M.\ Jacquin, M.\ Kudla, H.\ Piekarz,
                J.\ Piekarz, J.\ Pizzi, and J.\ Pniewski, Phys.\ Lett.\ B 
                {\bf 83}, 252 (1979).
\bibitem{dov90} C.~B.~Dover and H.~Feshbach, Ann.\ Phys.\ (N.Y.) 
                {\bf 198}, 321 (1990).
\bibitem{dov92} C.~B.~Dover and H.~Feshbach, Ann.\ Phys.\ (N.Y.) 
                {\bf 217}, 51 (1992).  
\bibitem{dovpr} C.\ B.\ Dover ({\it private communication}).
\bibitem{gib69} B.\ F.\ Gibson, A.\ Goldberg, and M.\ S.\ Weiss,
                Phys.\ Rev.\ {\bf 181}, 1486 (1969).
\bibitem{bod84} A.\ R.\ Bodmer, Q.\ N.\ Usmani, and J.\ A.\ Carlson,
                Nucl.\ Phys.\ {\bf A422}, 510 (1984).
\bibitem{ban82} H. Band\=o, Prog.\ Theor.\ Phys. {\bf 67}, 669 (1982).
\bibitem{bod66} A.\ R.\ Bodmer, Phys.\ Rev.\ {\bf 141}, 1387 (1966).
\bibitem{her67} R.\ C.\ Herndon, Y.\ C.\ Tang, Phys.\ Rev.\
                {\bf 153}, 1091 (1967); {\bf 159}, 853 (1967); {\bf 165},
                1093 (1969); R.\ H.\ Dalitz, R.\ C.\ Herndon, and Y.\
                C.\ Tang, Nucl.\ Phys.\ {\bf B47}, 109 (1972).
\bibitem{gib72} B.\ F.\ Gibson, A.\ Goldberg, and M.\ S.\
                Weiss, Phys.\ Rev.\ C {\bf 6}, 741 (1972).
\bibitem{gib72a} B.~F.~Gibson, A.~Goldberg, and M.~S.~Weiss,
                in {\it Few Particle Problems in Nuclear Interactions}, 
                (North Holland, Amsterdam, 1972), p.~188.
\bibitem{gal75} A.\ Gal, Adv.~Nucl.~Physics {\bf 8}, 1 (1975).
\bibitem{bod88} A.\ R.\ Bodmer and Q.\ N.\ Usmani, Nucl.\
                Phys.\ {\bf A477}, 621 (1988); Phys.\ Rev.\ C {\bf 31},
                1400 (1985); A.\ R.\ Bodmer, Q.\ N.\ Usmani, and J.\ A.\
                Carlson, Phys.\ Rev.\ C {\bf 29}, 684 (1984).
\bibitem{car91} J.\ A.\ Carlson, Proceedings of the {\it LAMPF Workshop on
                ($\pi,K$) Physics}, A.\ I.\ P.\ Conf.\ Proc.\ {\bf 224},
                ed.\ by B.\ F.\ Gibson, W.\ R.\ Gibbs, and M.\ B.\ Johnson,
                (American Institute of Physics, New York, 1991), pp.\ 198 -
                210.
\bibitem{Myi94} K.\ S.\ Myint and Y. Akaishi, Prog.\ Theo.\ Phys.\
                Supp.\ {\bf 117}, 251 (1994).
\bibitem{oka80} M.~Oka and K.~Yazaki, Phys.\ Lett.\ {\bf B90}, 
                41 (1980); M.~Oka and K.~Yazaki, Prog.\ Theor.\ Phys.\ 
                {\bf B66}, 556 (1981); 572 (1981).
\bibitem{fae82} A.~Faessler, F.~Fernandez, G.~L\={u}beck and 
                K.~Shimuzu, Phys.\ Lett.\ {\bf B112}, 201 (1982); 
                Nucl.\ Phys.\ {\bf A402}, 555 (1983).
\bibitem{oka84} M.~Oka and K.~Yazaki, {\em Quarks and Nuclei}, 
                ed.\ W.~Weise (World Scientific, Singapore, 1984),
                p.\ 489.
\bibitem{oka93} M.\ Oka, {\em Dibaryons in the Quark Model}, 
                Talk presented at the 10th International Symposium 
                (YAMADA conference XXXV) on High Energy Spin Physics, 
                1993.
\bibitem{koi90} Y.~Koike, K.~Shimizu and K.~Yazaki, Nucl.\ Phys.\ 
                {\bf A513}, 653 (1990).
\bibitem{oka87} M.~Oka, K.~Shimizu and K.~Yazaki, Nucl.\ Phys.\ 
                {\bf A464}, 700 (1987).   
\bibitem{dov84} C.~B.~Dover and A.~Gal, Prog.\ in Part.\ and Nucl.\ 
                Phys.\ {\bf 12}, 171 (1984).
\bibitem{gib94} B.~F.~Gibson, I.~R.~Afnan, J.~A.~Carlson and 
                D.~R.~Lehmann, Prog.\ Theor.\ Phys.\ Suppl.\ No.\ 
                117, 339 (1994).
\bibitem{rei68} R.~V.~Reid, Jr., Ann.\ Phys.\ (N.Y.) {\bf 50}, 
                411 (1968).
\bibitem{afn93} I.~R.~Afnan and B.~F.~Gibson, Phys.\ Rev.\ C {\bf 47}, 
                1000 (1993). 
\bibitem{yam54} Y.~Yamaguchi, Phys.\ Rev.\ {\bf 95}, 1628 (1954).

\bibitem{leh78} D.~R.~Lehman, M.~Rai and A.~Ghovanlou, Phys.\ Rev.\ C 
                {\bf 17}, 744 (1978)
\bibitem{leh74} A.~Ghovanlou and D.~R.~Lehman, Phys.\ Rev.\ C {\bf 9},
                1730 (1974).
\bibitem{esk92} A.~Eskandarian and I.~R.~Afnan, Phys. Rev. C {\bf 46}, 
                2344 (1992).
\bibitem{dov94} Carl~B.~Dover,  {\em Proceedings 
                of the U.S.-Japan Seminar on the Hyperon-Nucleon 
                Interaction}, ed.\ by B.~F.~Gibson, P.~D.~Barnes, and 
                K.~Nakai (World Scientific, Singapore, 1994), pp.\ 1-16.
\bibitem{bru54} K.~A.~Brueckner, C.~A.~Levinson amd H.~M.~Mahmoud, 
                Phys.\ Rev.\ {\bf 95}, 217 (1954).
\bibitem{gol57} J.~Goldstone, Proc.\ Roy.\ Soc.\ London {\bf A239}, 
                267 (1957).
\bibitem{day67} B.~D.~Day, Rev.\ Mod.\ Phys.\ {\bf 39}, 719 (1967).
\bibitem{yam85} Y.~Yamamoto and H.~Band\={o}, Prog.\ Theor.\ Phys.\ 
                {\bf 73}, 905 (1985).
\bibitem{kur85} Y.~Kurihara, Y.~Akaishi and H.~Tanaka, Phys.\ Rev.\ 
                C {\bf 31}, 971 (1985).
\bibitem{mon63} A.~S.~Mondal {\it et al.}, Nuovo Cim.\ {\bf 29}, 
                1 (1963).
\bibitem{dov83} C.~B.~Dover and A.~Gal, Ann.\ Phys.\ (N.Y.) {\bf 146}, 
                309 (1983).
\bibitem{alt67} E.~O.~Alt, P.~Grassberger and W.~Sandhas, Nucl.\ 
                Phys. {\bf B2}, 167 (1967).
\bibitem{afn89} I.~R.~Afnan and B.~F.~Gibson, Phys.\ Rev.\ C 
                {\bf40}, R7 (1989); {\bf 41}, 2787 (1990). 
\bibitem{afn74} I.~R.~Afnan and A.~W.~Thomas, Phys.\ Rev.\ C 
                {\bf 10}, 109 (1974).
\bibitem{bod87} A.\ R.\ Bodmer and Q.\ N.\ Usmani, Nucl.\ Phys.\
                {\bf A468}, 653 (1987).
\bibitem{usm95} A.~A.~Usmani, Phys.\ Rev.\ C {\bf 52}, 1773 (1995).

\end{references}
\end{document}